\RequirePackage{fix-cm}
\documentclass[natbib]{svjour3} 
\smartqed 
\usepackage{graphicx}
\usepackage{amsmath}
\usepackage{amssymb}
\usepackage{gensymb} 
\usepackage[utf8]{inputenc} 
\journalname{Celestial Mechanics and Dynamical Astronomy}
\begin{document}

\title{Study and application of the resonant secular dynamics beyond Neptune}

\author{Melaine~Saillenfest$^{1,2}$ \and Marc~Fouchard$^{1,3}$ \and Giacomo~Tommei$^{2}$ \and Giovanni~B.~Valsecchi$^{4,5}$}
\authorrunning{Melaine Saillenfest \and Marc Fouchard \and Giacomo Tommei \and Giovanni B. Valsecchi} 

\institute{
   M. Saillenfest \at
   melaine.saillenfest@obspm.fr\\
   \\
   $^1$ IMCCE, Observatoire de Paris, 77 av. Denfert-Rochereau, 75014 Paris, France\newline
   $^2$ DM, Università di Pisa, Largo Bruno Pontecorvo 5, 56127 Pisa, Italia\newline
   $^3$ LAL-IMCCE, Université de Lille, 1 Impasse de l’Observatoire, 59000 Lille, France\newline
   $^4$ IAPS-INAF, via Fosso del Cavaliere 100, 00133 Roma, Italia\newline
   $^5$ IFAC-CNR, via Madonna del Piano 10, 50019 Sesto Fiorentino, Italia
}

\date{Received: 7 July 2016 / Revised: 24 September 2016 / Accepted: 26 September 2016}

\maketitle

\begin{abstract}
   We use a secular representation to describe the long-term dynamics of transneptunian objects in mean-motion resonance with Neptune. The model applied is thoroughly described in \cite{SAILLENFEST-etal_2016}. The parameter space is systematically explored, showing that the secular trajectories depend little on the resonance order. High-amplitude oscillations of the perihelion distance are reported and localised in the space of the orbital parameters. In particular, we show that a large perihelion distance is not a sufficient criterion to declare that an object is detached from the planets. Such a mechanism, though, is found unable to explain the orbits of Sedna or $2012\text{VP}_{113}$, which are insufficiently inclined (considering their high perihelion distance) to be possibly driven by such a resonant dynamics. The secular representation highlights the existence of a high-perihelion accumulation zone due to resonances of type $1\!:\!k$ with Neptune. That region is found to be located roughly at $a\in[100;300]$~AU, $q\in [50;70]$~AU and $I\in [30;50]\degree$. In addition to the flux of objects directly coming from the Scattered Disc, numerical simulations show that the Oort Cloud is also a substantial source for such objects. Naturally, as that mechanism relies on fragile captures in high-order resonances, our conclusions break down in the case of a significant external perturber. The detection of such a reservoir could thus be an observational constraint to probe the external Solar System.
      
\end{abstract}

\section{Introduction}
   Mean-motion resonances with Neptune are a well-known mechanism responsible for large orbital variations of transneptunian objects. Since that statement relies almost only on the tracking of simulated samples \citep{HOLMAN-WISDOM_1993, DUNCAN-etal_1995, GLADMAN-etal_2002, GOMES-etal_2005}, it is yet unclear what kind of secular trajectories can actually be adopted by the resonant particles. Indeed, the resonance captures being a relatively rare process, we cannot rely only on numerical integrations to efficiently explore the space of possible behaviours. Furthermore, because of our current lack of orbital data in the intermediate region between the Kuiper Belt and the Oort Cloud, numerous hard-to-test hypotheses are presented in the literature to investigate its distribution. That interest was renewed after the study by \cite{BATYGIN-BROWN_2016} about the possible existence of a ninth planet beyond the Kuiper Belt. In that context, it would be useful to have a general picture of the resonant dynamics in the external Solar System, including the most extreme (even if improbable?) orbits.
   
   Being the resonant trajectories beyond Neptune quasi-integrable, they can be described by secular theories. As such, \cite{KOZAI_1985} adapted his secular model from \cite{KOZAI_1962} to take into account a mean-motion resonance. He applied it in particular to Pluto. Afterwards, that model was improved and used more generally in the transneptunian region \citep{GOMES-etal_2005,GOMES_2011,GALLARDO-etal_2012}. However, even the improved versions do not represent accurately the secular trajectories of the resonant particles (or they are cumbersome to use), because they require a prior knowledge of the evolution of one degree of freedom (namely the resonant angle $\sigma$ and the semi-major axis $a$). In a first paper \citep[][hereafter called \emph{Paper I}]{SAILLENFEST-etal_2016}, we developed a new resonant secular model, designed to bypass these limitations by the use of the adiabatic invariant hypothesis. For a given set of internal planets on circular and coplanar orbits, it provides a very accurate one-degree-of-freedom representation of the resonant secular trajectories (self-consistency). As stated by \cite{THOMAS-MORBIDELLI_1996}, such a model can be seen as the dominant term of an expansion in powers of the planetary eccentricities and inclinations. It can thus be used to perform a global analysis of the resonant trajectories beyond Neptune for a dynamics solely driven by the known planets.
   
   The first goal of this paper is to study the different kinds of dynamics allowed by a resonant interaction with Neptune in the far Solar System. As anticipated by previous studies, there is a vast panel of possible interesting trajectories which have to be organised and characterised with regard to the orbital parameters. This can be seen as a cold analysis of the semi-analytical mathematical model described in Paper I. Secondly, we want to determine to what extend such a model can apply to the known distant transneptunian objects. The current data are scarce and the resonance captures are hard to detect, however, some very general arguments can be used to select the potentially interesting bodies for this study. Finally, it is crucial to establish if these results could imply some selection in the distribution of the transneptunian objects, such as accumulation regions in the space of the orbital elements. Indeed, the confrontation with future observations is the only way to test our knowledge of the distant Solar System.
   
   In Sect.~\ref{sec:model}, we recall briefly the resonant secular model developed in Paper I and applied throughout this paper. In Sect.~\ref{sec:explor}, we describe the variety of trajectories allowed by a resonant interaction with Neptune with respect to the fixed parameters of the model. In particular, there is a topological difference between the resonances of type $1\!:\!k$ (Sect.~\ref{subsec:type1k}) and other types of resonances (Sect.~\ref{subsec:typeGen}), but no major change of geometry is observed when varying the resonance order. A particular family of periodic secular orbits is highlighted, allowing to bypass the discontinuity line inherent to resonances of type $1\!:\!k$ (Sect.~\ref{subsec:playDisc}). The parameter $J$, related to the amplitude of oscillations of the resonant angle, is studied separately (Sect.~\ref{subsec:highAmp}). Then, Sect.~\ref{sec:app} presents the ranges of parameters corresponding to ``interesting" geometries in the phase portraits, that is allowing large variations of the perihelion distance and/or the libration of the argument of perihelion. The known distant transneptunian objects are localised with respect to these regions in order to determine whether a resonance capture could affect significantly their dynamics. Section~\ref{sec:trapping} describes a mechanism of high-perihelion capture associated to resonances of type $1\!:\!k$. It predicts an accumulation zone in the space of orbital elements and complements the results by~\cite{GOMES-etal_2005}. Finally, we show in Sect.~\ref{sec:OC} that the Oort Cloud is an effective source of Scattered Disc objects and thus contributes to feed the accumulation zone.
   
\section{The secular Hamiltonian function}\label{sec:model}

   As in \cite{KOZAI_1985} or \cite{THOMAS-MORBIDELLI_1996}, we use a set of $N$ planets evolving on circular and coplanar orbits. In Delaunay heliocentric elements $(L,G,H,\ell,g,h)$, the Hamiltonian function for a test-particle is then:
   \begin{equation}\label{eq:Hgen}
      \mathcal{H}\Big(\{\Lambda_i\},L,G,H,\{\lambda_i\},\ell,g,h\Big) = \mathcal{H}_0\Big(\{\Lambda_i\},L\Big) + \varepsilon\,\mathcal{H}_1\Big(L,G,H,\{\lambda_i\},\ell,g,h\Big)
   \end{equation}
   where the integrable part and the perturbation write respectively:
   \begin{equation}\label{eq:H0eH1}
      \left\{
      \begin{aligned}
         \mathcal{H}_0 &= -\frac{\mu^2}{2L^2} + \sum_{i=1}^{N}n_i\,\Lambda_i \\
         \varepsilon\,\mathcal{H}_1 &= - \sum_{i=1}^{N}\mu_i\left(\frac{1}{|\mathbf{r}-\mathbf{r}_i|} - \mathbf{r}\cdot\frac{\mathbf{r}_i}{|\mathbf{r}_i|^3}\right)
      \end{aligned}
      \right.
   \end{equation}
   In these expressions, $\mathbf{r}$ and $\mathbf{r}_i$ are the heliocentric positions of the particle and of the $i$th planet, and $\mu$ and $\mu_i$ are the gravitational parameters of the Sun and of the $i$th planet, respectively. The momenta $\{\Lambda_i\}$ are conjugated to the mean longitudes $\{\lambda_i\}$ of the planets. They allow the definition of an autonomous Hamiltonian, the constants $\{n_i\}$ being the mean motions of the planets. In the Hamiltonian~\eqref{eq:H0eH1}, we have then $\mathbf{r}_i \equiv \mathbf{r}_i(\lambda_i)$ and $\mathbf{r} \equiv \mathbf{r}(L,G,H,\ell,g,h)$. Of course, the Delaunay canonical coordinates are directly linked to the usual Keplerian elements $(a,e,I,\omega,\Omega,M)$ of the particle.
   
   Let us consider a single resonance of type $k_p\!:\!k$ with a resonant angle of the form:
   \begin{equation}
      \sigma = k\,\lambda - k_p\,\lambda_p - (k-k_p)\,\varpi
      \text{\ \ \ ,\ \ \ } k,k_p\in\mathbb{N}\text{\ \ \ ,\ \ \ }k>k_p
   \end{equation}
   where $\lambda$ and $\lambda_p$ are the mean longitudes of the particle and of the planet $p$ involved, and $\varpi=\omega+\Omega$. The integer $k-k_p$ is traditionally called the \emph{resonance order}. In our case, it is directly linked to the magnitude of the semi-major axis $a$ of the particle. In order to study the dynamics inside or around the $k_p\!:\!k$ resonance, we define $\sigma$ as a new canonical coordinate \citep[as for instance in][]{MILANI-BACCILI_1998}. This is done by a linear transformation applied to the angles:
   \begin{equation}
      \begin{pmatrix}
         \sigma \\
         \gamma \\
         u \\
         v
      \end{pmatrix}
      =
      A
      \begin{pmatrix}
         \ell \\
         \lambda_p \\
         g \\
         h
      \end{pmatrix}
      =
      \begin{pmatrix}
         k & -k_p & k_p & k_p \\
         c & -c_p & c_p & c_p \\
         0 & 0 & 1 & 0 \\
         0 & 0 & 0 & 1
      \end{pmatrix}
      \begin{pmatrix}
         \ell \\
         \lambda_p \\
         g \\
         h
      \end{pmatrix}
   \end{equation}
   where $c$ and $c_p$ are integer coefficients chosen such that $\det(A) = 1$. The coordinates $(\Sigma,\Gamma,U,V)$ are then obtained by applying $(A^T)^{-1}$ on the conjugated momenta, and the variables $\{\lambda_{i\neq p}\}$ and $\{\Lambda_{i\neq p}\}$ remain unchanged. Assuming that $\sigma$ librates slowly, we have now three different time-scales:
   \begin{itemize}
      \item[$\bullet$] the short periods ($\gamma$ and $\{\lambda_{i\neq p}\}$)
      \item[$\bullet$] the semi-secular periods (oscillation of the resonant angle $\sigma$)
      \item[$\bullet$] the secular periods (precession of $u=\omega$ and $v=\Omega$)
   \end{itemize}
   If the dynamics does not present any short time-scale chaos, we can get rid of the short-period terms by a near-identity canonical transformation. At first order of the planetary masses, the Hamiltonian function in the new coordinates writes then (see Paper I):
   \begin{equation}
      \mathcal{K}\big(\Sigma,U,V,\sigma,u\big) = \mathcal{K}_0\big(\Sigma\big) + \varepsilon\,\mathcal{K}_1\big(\Sigma,U,V,\sigma,u\big)
   \end{equation}
   where:
   \begin{equation}
      \mathcal{K}_0 = -\frac{\mu^2}{2\,(k\Sigma)^2} - n_pk_p\Sigma
   \end{equation}
   and where $\varepsilon\,\mathcal{K}_1$ is obtained by computing numerically the average of $\varepsilon\,\mathcal{H}_1$ with respect to the independent angles $\gamma$ and $\{\lambda_{i\neq p}\}$. Thanks to the cylindrical symmetry implied by the circular and coplanar orbits of the planets, the angle $v=\Omega$ has disappeared, making $V$ a semi-secular constant of motion:
   \begin{equation}\label{eq:Vconst}
      V = \sqrt{\mu a} \left(\eta - k_p/k\right) = const
      \hspace{0.5cm}\text{with}\hspace{0.5cm}
      \eta = \sqrt{1-e^2}\cos I
   \end{equation}
   The momentum $\Gamma$ is also a semi-secular constant and it is conveniently chosen equal to $0$. We are left with a two-degree-of-freedom system, but with a clear hierarchy between the two associated time-scales. Hence, we can use the adiabatic invariant hypothesis \citep[as described in][]{HENRARD_1993} to reduce the problem to a one-degree-of-freedom system. The secular Hamiltonian function writes finally:
   \begin{equation}\label{eq:Fgen}
      \mathcal{F}\big(J,U,V,u\big) = \mathcal{K}(\Sigma_0,U,V,\sigma_0,u)
   \end{equation}
   where $(\Sigma_0,\sigma_0)$ represents any point of the level curve of $\mathcal{K}$ for $(U,u)$ fixed which encloses the signed area $2\pi J$. The momentum $J$ is a secular constant of motion, related to the oscillation amplitude of $\Sigma=\sqrt{\mu a}/k$ and $\sigma$. In particular if $J=0$, the coordinates $(\Sigma,\sigma)$ are fixed at the libration centre of the resonance and undergo only a secular drift due to the variation of $(U,u)$. The specific level curve of $\mathcal{K}$ in the $(\Sigma,\sigma)$ plane is obtained by numerical integrations and a Newton algorithm. The method of \cite{HENON_1982} is used to stop the integrations exactly after a complete cycle. Since the secular Hamiltonian function~\eqref{eq:Fgen} has only one degree of freedom, the secular dynamics can be described by plotting its level curves in the plane $(U,u)$ with $V$ and $J$ as parameters. We recall that $u=\omega$ and $U = \sqrt{\mu a} \left(\sqrt{1-e^2} - k_p/k\right)$, and that $\mathcal{F}$ is $\pi$-periodic with respect to $u$.
   
   The experience shows that the secular variations of the semi-major axis $a$ are always rather small, such that it is never far from a central approximate value $a_0$. In order to get a more direct representation, we will thus replace the constant $V$ (Eq.~\ref{eq:Vconst}) by the parameter:
   \begin{equation}
      \eta_0 = \frac{V}{\sqrt{\mu a_0}} + \frac{k_p}{k}
   \end{equation}
   In that expression, the value of the ``reference semi-major axis" $a_0$ is a simple matter of definition and acts only as a short-cut from $V$ to an approximation of the corresponding secular Keplerian elements. In the same way, the variable $U$ can be replaced by a reference perihelion $\tilde{q}$, directly linked to $U$ through $a_0$ via a reference eccentricity $\tilde{e}$ (see Paper I). The plane $(\tilde{q},\omega)$ is thus entirely equivalent to the plane $(U,u)$, and the parameter $\eta_0$ is entirely equivalent to the $V$ constant. Finally, we can also define a reference inclination by $\cos\tilde{I} = \eta_0/\sqrt{1-\tilde{e}\,^2}$, so that the parameter $\eta_0$ acts as the Kozai constant of the non-resonant case, linking the inclination and the eccentricity (even if this time, it is in an approximative way). In the following, this model is applied to the four giant planets, the mass of the inner ones being added to the Sun. 
   
\section{Exploration of the parameter space}\label{sec:explor}
   To prevent any confusion between the three time-scales involved, note that in the following, the terms ``resonance island" always refer to the semi-secular Hamiltonian $\mathcal{K}$ for $(U,u)$ fixed, that is to the usual oscillations of the semi-major axis and of the resonance angle $\sigma$. Regarding the equilibrium points of the \emph{secular} Hamiltonian $\mathcal{F}$, that is in the plane $(\omega,\tilde{q})$, we will speak generically of ``libration islands" (because strictly speaking they are not resonant even if some authors call them ``Kozai resonances").

   In order to determine the influence of the chosen resonance and of the parameters on the phase space, we plotted a vast collection of level curves for various resonances with semi-major axes between $80$ and $600$ AU. In that section, we describe our results and develop a general picture of the resonant secular dynamics beyond Neptune ($q>a_\text{N}$). We will see that the geometry of the phase portraits depends mostly of what we call the ``resonance type", that is the coefficient $k_p$ involved ($1,2,3...$). Note that even very high-order resonances can present interesting geometries with wide libration zones. The corresponding probability of capture and stability are of course lower, but these considerations are not studied in this section: here we just suppose that the particle \emph{is} trapped in the chosen resonance and we describe the secular effects it would produce.
   
   Whatever the resonance considered, we found that for $\eta_0 \approx \pm 1$ or $\eta_0 \approx 0$, that is for orbits nearly circular-coplanar or perpendicular to the planetary plane\footnote{A parameter $\eta_0=0$ is also attainable for $e=1$ (whatever the inclination), but since we are interested in perihelion distances always beyond Neptune, we will not deal with parabolic orbits in this paper.}, the level curves of the secular Hamiltonian are very ``flat". In these cases, the resonant dynamics is thus very similar to the generic non-resonant one: circulation of $\omega$ with very small oscillations of $\tilde{q}$. In particular, note that the upper features on Fig.~11e by \cite{GALLARDO-etal_2012} are irrelevant\footnote{A fixed libration centre is very unsuitable in this region because it varies actually between $0$ and $2\pi$. That comment holds also for their Figs.~11f-h beyond about $50$~AU. In the lower part, on the contrary, a resonance island centred around $60\degree$ does exist (even if it actually shifts and deforms a bit).}.
   
   On the other hand, there is always a range of $\eta_0$ both for retrograde ($\eta_0<0$) or prograde ($\eta_0>0$) orbits, for which the secular Hamiltonian shows equilibrium points for $\omega$ and $\tilde{q}$. In the following, we will refer to that interval of $\eta_0$ by the \emph{range of interest} because it can allow wide perihelion variations and/or confinement regions for $\omega$.
   
   Additionally, the classic non-resonant Kozai islands show up where the resonant part of the Hamiltonian function weakens\footnote{Strictly speaking, the non-resonant Hamiltonian is defined with the semi-major axis as a constant parameter. However, the resonant interaction does not cause $a$ to vary enough for the Kozai islands to be notably distorted.}. This happens if their typical inclinations (near $63\degree$ or $117\degree$) correspond to a sufficiently small eccentricity, that is for a parameter $\eta_0$ far enough from $0$. In the intermediate regime where the non-resonant and resonant parts have comparable strength, the interaction of these islands with the purely resonant features produces complex geometries.
   
   \subsection{Single resonance island and $J$ near zero}\label{subsec:typeGen}
\begin{figure}
      \centering
      \includegraphics[width=\textwidth]{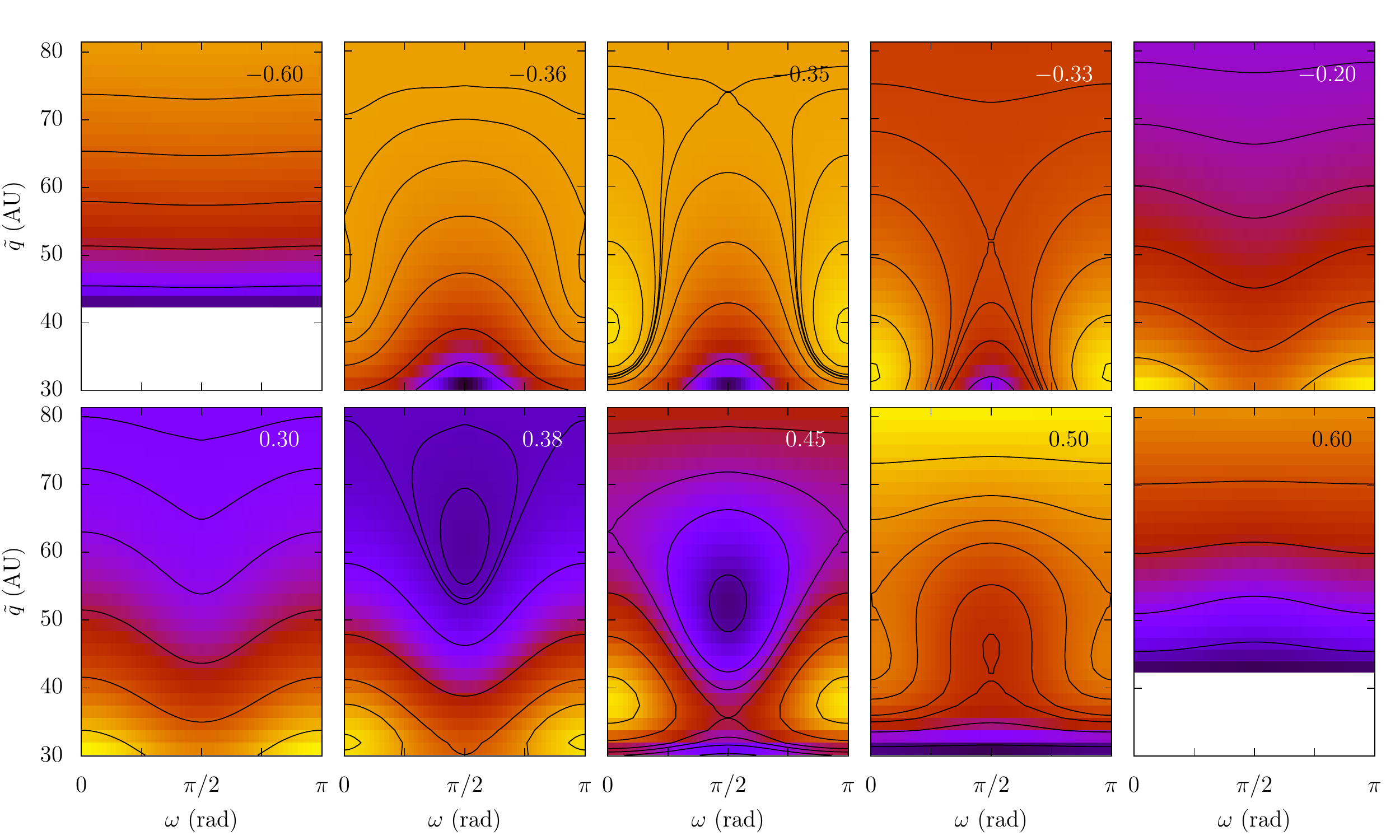}
      \caption{Typical dependence on the parameter $\eta_0$ for orbits in resonances of type $k_p\neq 1$ with Neptune. The resonance taken in example here is the $2\!:\!37$ (reference semi-major axis chosen $a_0=210.9911$~AU). All of these graphs are plotted for $J=0$ and $\eta_0$ is indicated in the upper right corners. On the Y-axis of each graph, the reference inclinations can be obtained by $\cos \tilde{I}=\eta_0/\small\sqrt{1-(1-\tilde{q}/a_0)^2}$. The white regions on the first and last graphs are unreachable with these values of $\eta_0$ (they would require a cosine of inclination lower than $-1$ or higher than $1$, respectively).}
      \label{fig:Hsec_general}
      \vspace{0.5cm}
      \includegraphics[width=\textwidth]{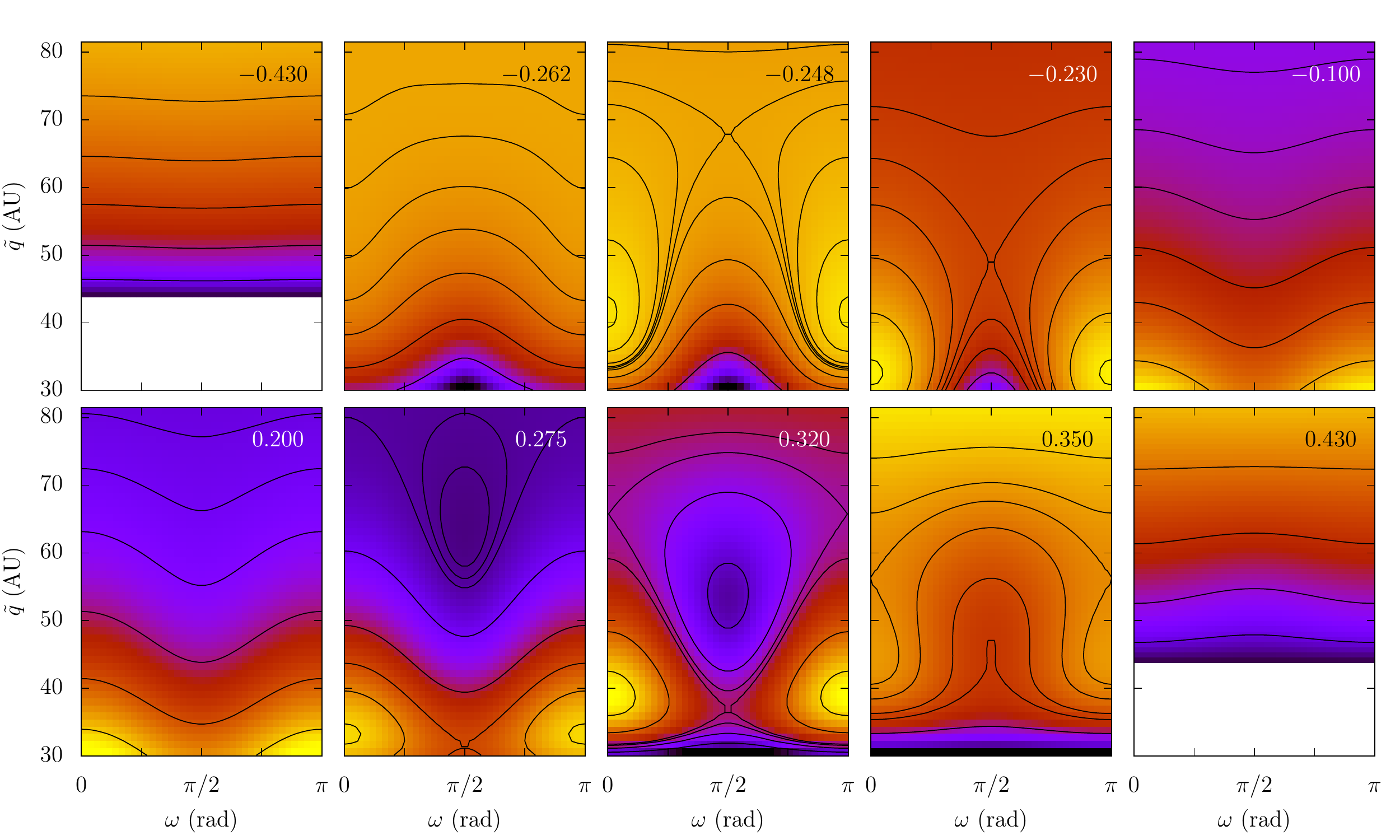}
      \caption{Same as Fig.~\ref{fig:Hsec_general} for the resonance $2\!:\!115$ (reference semi-major axis chosen $a_0=449.3602$~AU). The parameter $\eta_0$ has been tuned to give approximatively the same phase portraits as in Fig.~\ref{fig:Hsec_general}.}
      \label{fig:Hsec_general2}
   \end{figure}

\begin{figure}
      \centering
      \includegraphics[width=\textwidth]{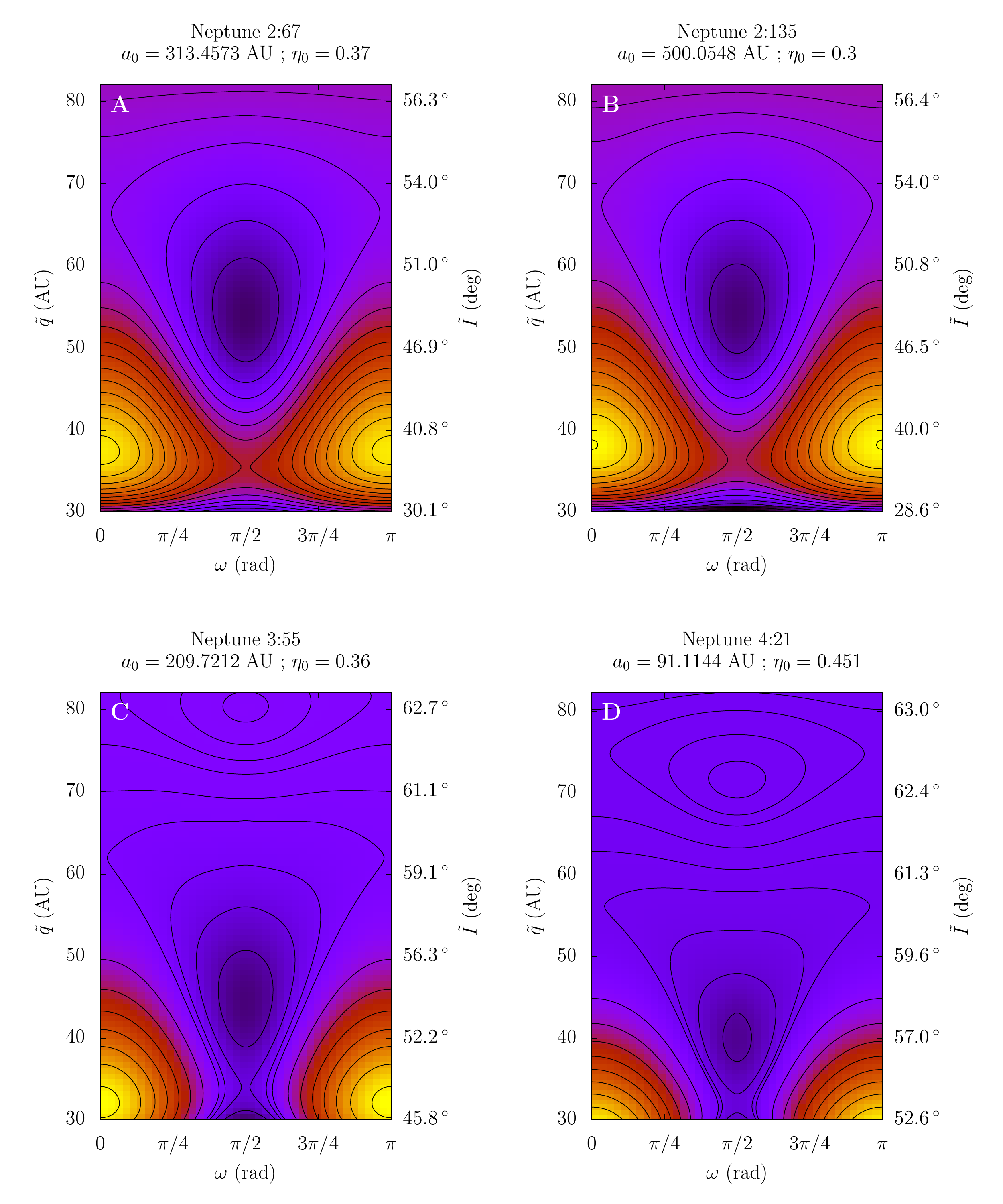}
      \caption{Typical geometries for prograde orbits in resonances of type $k_p\neq 1$ with Neptune, with a parameter $J=0$. The resonances and corresponding $a_0$ are indicated above the graphs, as well as the parameter $\eta_0$ used. A large variety of semi-major axes are presented to stress that the secular phase space depends little on the resonance order. The graphs A and B are very similar since they have both $k_p=2$, but correspond to different values of $\eta_0$. On the graph C ($3\!:\!k$ resonance), the two islands still coexist but a slight increase of $\eta_0$, shifting up the $\omega=0$ island at a perihelion similar to the graphs A and B, would make disappear the island at $\omega=\pi/2$. On the graph D ($4\!:\!k$ resonance), the coexistence is even impossible: an increase of $\eta_0$ would make disappear the $\omega=\pi/2$ island before the rise of the $\omega=0$ one. On the graphs C and D, the classic Kozai island is visible at high inclinations.}
      \label{fig:prograde}
   \end{figure}

\begin{figure}
      \centering
      \includegraphics[width=\textwidth]{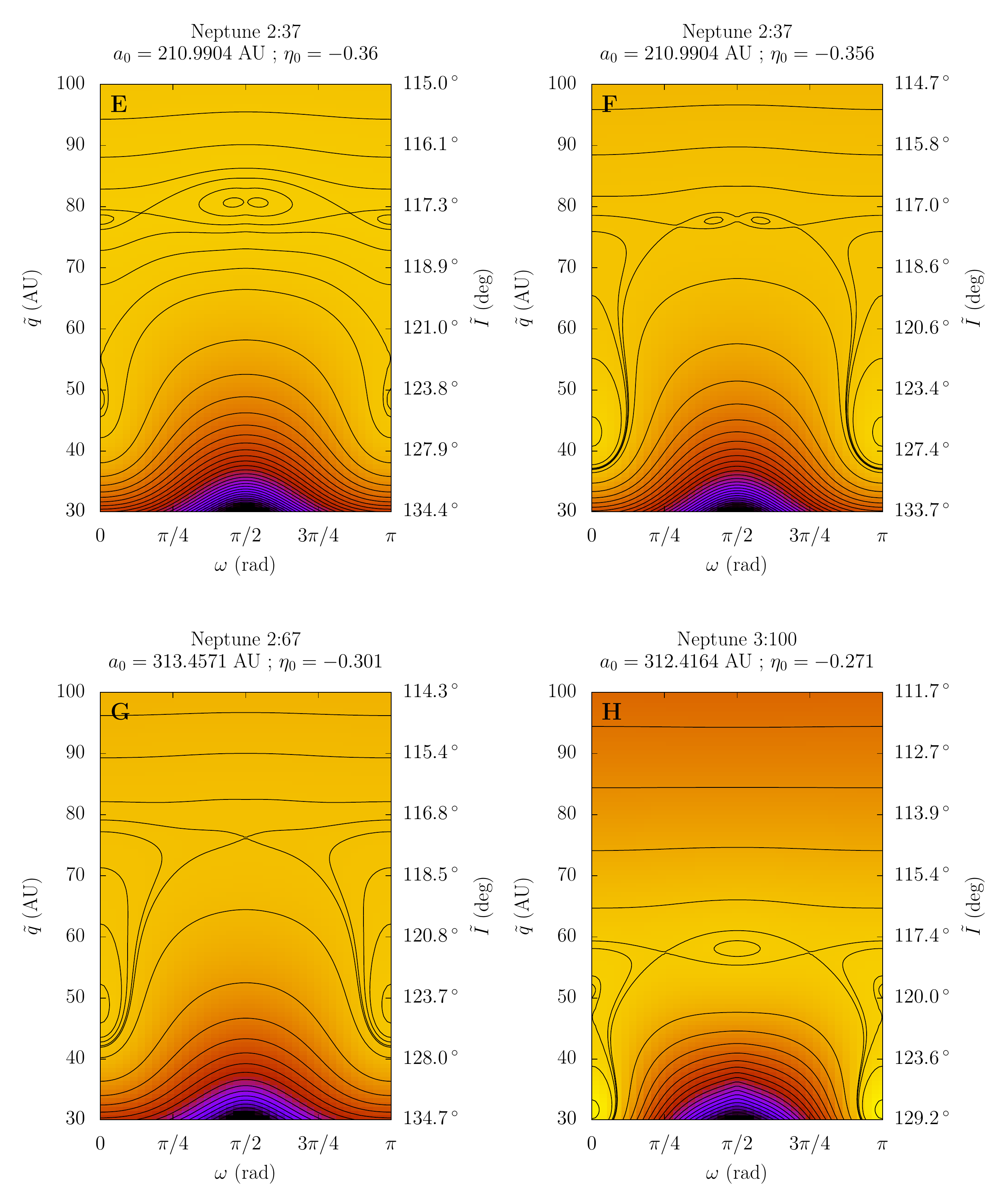}
      \caption{Same as Fig.~\ref{fig:prograde} for retrograde orbits. The graphs E and F present the same resonance for a slightly different parameter $\eta_0$. Passing in the neighbourhood of the $\omega=0$ island, the classic Kozai island (at about $117\degree$) deforms and merges to form high-amplitude oscillation zones for the perihelion distance. That enlargement is also clearly visible on the graphs G and H, which present two resonances with neighbour semi-major axes but different types ($2\!:\!k$ and $3\!:\!k$). As for prograde orbits, $k_p$ modifies the respective ranges of $\eta_0$ for the appearance of the different features, which gives rise to different geometries.}
      \label{fig:retrograde}
   \end{figure}

   Figures~\ref{fig:Hsec_general} and \ref{fig:Hsec_general2} show the typical evolution of the phase portraits for $k_p\neq 1$ with respect to the parameter $\eta_0$. As explained in Paper I, such resonances present a single resonance island for any value of the parameters\footnote{Some resonances other than $1\!:\!k$ can actually present two resonance islands, but the corresponding ranges of parameters for $q>a_\text{N}$ are so narrow that we will dismiss that case in the present paper.} so the corresponding secular phase space is devoid of discontinuity line. Figures~\ref{fig:prograde} and \ref{fig:retrograde} show further details and comparisons between different resonances of type $k_p\neq 1$ for $\eta_0$ in the range of interest. We can make the following general observations:
   \begin{itemize}
      \item[$\bullet$] Either for prograde or retrograde orbits, some range of $\eta_0$ allows a libration island at $\omega=0$. When $\eta_0$ tends to $0$ (orbit perpendicular to the planetary plane), that island gets closer the orbit of Neptune and disappears below it.
      \item[$\bullet$] For prograde orbits, a range of $\eta_0$ allows an additional island at $\omega=\pi/2$.
      \item[$\bullet$] Two resonances of the same ``type" (that is with the same coefficient $k_p$) present very similar geometries, but located in a different range of $\eta_0$ (compare Figs.~\ref{fig:Hsec_general} and \ref{fig:Hsec_general2}). Since the features are located both at the same $\tilde{q}$ and $\tilde{I}$, the parameters $\eta_0$ giving the same geometries for two different resonances are the ones implying approximatively the same interval of $\tilde{I}$ inside the same interval of $\tilde{q}$ (see the graphs A and B of Fig.~\ref{fig:prograde}).
      \item[$\bullet$] For resonances of type $2\!:\!k$, the two respective ranges of $\eta_0$ for the existence of the equilibrium points at $\omega =0$ and $\pi/2$ are rather the same, so the islands can coexist on the same phase portrait (graphs A and B of Fig.~\ref{fig:prograde}). For $k_p=3$ and beyond, the $\omega=\pi/2$ island appears at higher inclinations, for which the $\omega=0$ island is much lower (graph C) or even inside the orbit of Neptune (graph D).
      \item[$\bullet$] On some graphs of Figs.~\ref{fig:prograde} and Fig.~\ref{fig:retrograde}, the classic non-resonant Kozai island is clearly visible. It can either remain isolated (graphs C, D) or interact with the resonant features, to create new islands (graphs E, H at $\omega=0$ and E, F on both sides of $\omega=\pi/2$) or enlarge the existing ones (graph F, G, H).
      \item[$\bullet$] Contrary to the geometry of the phase portraits, the time-scale highly depends on the coefficient $k$, that is of the semi-major axis of the particle (at $k_p$ constant). As an example, numerical integrations of the averaged system show that the biggest loop of the graph A is completed in about $8$~Gyrs, whereas the analogous loop takes $16$~Gyrs in the graph B.
   \end{itemize}
   
   \subsection{Resonances of type $1\!:\!k$ for near-zero values of $J$}\label{subsec:type1k}
\begin{figure}
      \centering
      \includegraphics[width=\textwidth]{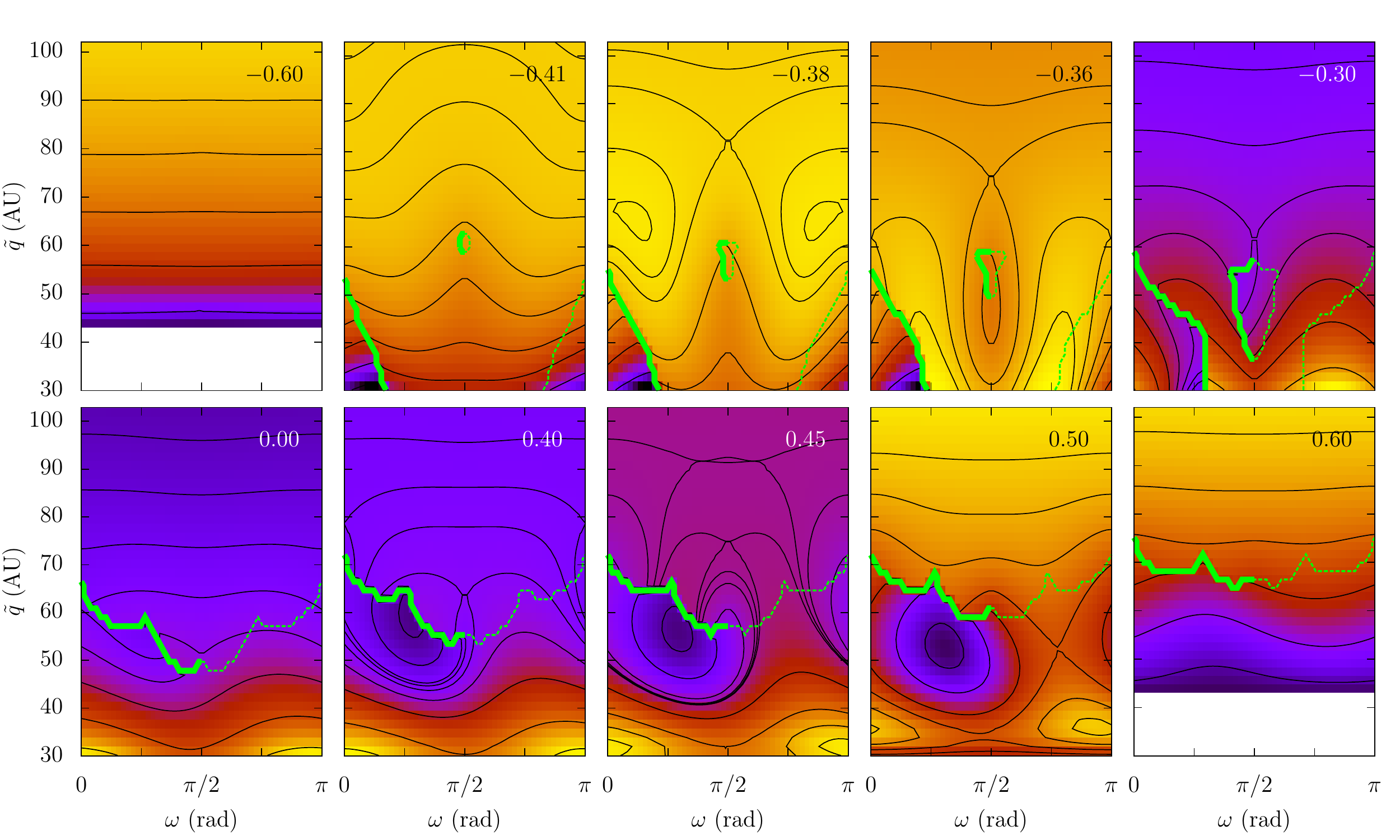}
      \caption{Typical dependence on the parameter $\eta_0$ for orbits in resonances of type $1\!:\!k$ with Neptune. The resonance taken in example here is the $1\!:\!19$ (reference semi-major axis chosen $a_0=214.7763$~AU). When there is a change of topology in the semi-secular phase space (from a two-resonance-island to a one-resonance-island configuration), the secular phase portrait can either present a discontinuity (thick green line) or a soft transition (thin dashed green line). In the regions where there are two resonance islands (below the green line or inside it when it is closed), we chose $\sigma$ to oscillate inside the right one. All of these graphs are plotted for $J=0$ and $\eta_0$ is indicated in the upper right corners. On the Y-axis of each graph, the reference inclinations can be obtained by $\cos \tilde{I}=\eta_0/\small\sqrt{1-(1-\tilde{q}/a_0)^2}$. The white regions on the first and last graphs are unreachable with these values of $\eta_0$ (they would require a cosine of inclination lower than $-1$ or higher than $1$, respectively).}
      \label{fig:Hsec_general_1k}
   \end{figure}

\begin{figure}
      \centering
      \includegraphics[width=\textwidth]{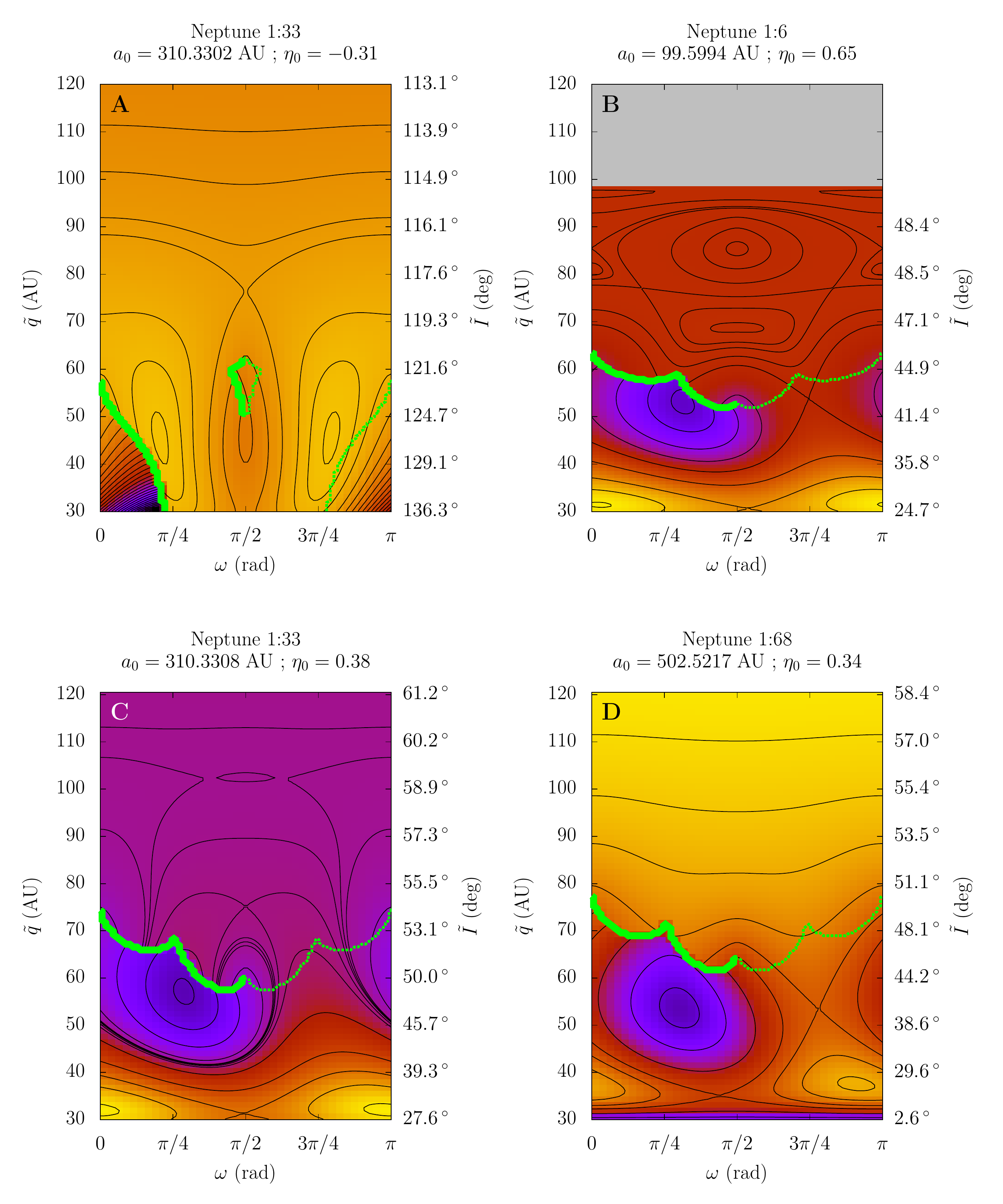}
      \caption{Typical geometries for resonances of type $1\!:\!k$ with Neptune, with a parameter $J=0$ and oscillations of $\sigma$ inside the right island. The resonances and corresponding $a_0$ are indicated above the graphs, as well as the parameter $\eta_0$ used. These graphs have to be compared and located in Fig.~\ref{fig:Hsec_general_1k}: note the similar geometries when changing $k$ but for a different scale of the parameter $\eta_0$. The graph B shows that for small semi-major axes, the geometry at high perihelion distance is modified by the proximity of the circular orbit. The graph D shows that these trajectories are not restricted to high-inclination regimes and that they can occasion very wide variations of the inclination.}
      \label{fig:pro-retro1k}
   \end{figure}

   Figure~\ref{fig:Hsec_general_1k} presents the typical evolution of the phase portraits for $k_p=1$ with respect to the parameter $\eta_0$. As that kind of resonances can present two resonance islands in definite regions of the plane $(\omega,\tilde{q})$, we chose to describe the case of $\sigma$ oscillating inside the ``right" one\footnote{As the semi-secular Hamiltonian $\mathcal{K}$ is symmetric in $\omega$ with respect to $\pi/2$, the secular level curves for oscillations inside the left island are obtained by the transformation $\omega\rightarrow\pi-\omega$.}. The green line divides the zones associated with two resonance islands from the zones associated with a single one. If it is crossed, there can be either a discontinuity (if the vanishing island is the very one occupied by the particle), or a soft transition. For particles  following a level curve leading to the discontinuity, another secular model is necessary after the crossing, with a different parameter $J$. Naturally, the discontinuity is only one-way: if the particle comes \emph{from} the one-island side (for prograde orbits, this means from the high-perihelion region), the appearance of the second island does not imply any particular transition ($J$ is continuously conserved). The appropriate secular representation should though be used ($\sigma$ oscillating in the left or right resonance island). Figure~\ref{fig:pro-retro1k} shows further details and comparisons between different resonances of type $1\!:\!k$ for $\eta_0$ in the range of interest. Our general conclusions are listed below:
   \begin{itemize}
      \item[$\bullet$] For retrograde orbits, there is only one resonance island almost everywhere in the plane $(\omega,\tilde{q})$, which produces symmetric level curves with respect to $\pi/2$ (see Fig.~\ref{fig:Hsec_general_1k}, upper graphs). The small regions with two resonance islands are located in a small range of perihelion at $\omega=\pi/2$ and near the collision points with Neptune ($\omega=0$ or $\pi$ and $q=a_\text{N}$). When $\eta_0$ tends to $0$, these three regions merge and eventually occupy all the bottom part of the graphs (as for prograde orbits).
      \item[$\bullet$] For prograde orbits, the geometries are actually very similar to the ones obtained for $k_p=2$, apart from the asymmetry and the discontinuity line induced by the two resonance islands. In other words, a range of $\eta_0$ allows the same libration islands at $\omega=0$ and $\pi/2$, which are however shifted and more or less distorted. The $\omega=\pi/2$ island can be besides truncated by the discontinuity line. As before, an increase of $\eta_0$ simultaneously shifts up the $\omega=0$ equilibrium point and down the $\pi/2$ one until they both disappear. The $\pi/2$ island is the last to vanish.
      \item[$\bullet$] For retrograde orbits, a range of $\eta_0$ allows an equilibrium point at $\omega=0$. When $\eta_0$ tends to zero, that equilibrium shifts toward the orbit of Neptune, but contrary to resonances of type $k_p\neq 1$, it splits in two (see Fig.~\ref{fig:Hsec_general_1k} for $\eta_0=-0.38$). This allows to partially avoid the discontinuity curve (Fig.~\ref{fig:pro-retro1k}, graph A). During this process, a small range of $\eta_0$ allows an additional equilibrium point at $\omega=\pi/2$, but the associated libration island is truncated by the discontinuity curve (see Fig.~\ref{fig:Hsec_general_1k} for $\eta_0=-0.36$ and Fig.~\ref{fig:pro-retro1k}, graph A).
      \item[$\bullet$] As for other kinds of resonances, the interaction with the classic non-resonant Kozai island can enlarge the resonant features to create spectacular excursions of the perihelion distance. On the graph C of Fig.~\ref{fig:pro-retro1k}, for instance, the classic Kozai island is located at $\tilde{q}=140$ AU (out of the plot), and produces a possible evolution from $\tilde{q}\approx 41$ AU to $107$ AU. A similar kind of enlargement is also visible on Fig.~\ref{fig:Hsec_general_1k} for $\eta_0=0.45$.
      \item[$\bullet$] For small semi-major axes (say $<130$ AU), the geometries at high perihelion distances can be more complex because of the proximity of the circular orbit. This can create various new equilibrium points, as on the graph B of Fig.~\ref{fig:pro-retro1k} (four additional islands). A different behaviour for small semi-major axes was also reported in the non-resonant case \citep[see][or Paper I]{GALLARDO-etal_2012}.
   \end{itemize}
   
   \subsection{Playing with the secular discontinuity line}\label{subsec:playDisc}
   From Sect.~\ref{subsec:type1k}, we know that any secular representation for prograde resonances of type $1\!:\!k$ presents a one-way discontinuity line in the plane $(\omega,\tilde{q})$ from low to high perihelion distances. Since that line is only half-width\footnote{The discontinuity line spans in $\omega\in [0;\pi/2]$ or $[\pi/2;\pi]$ if the particle occupies the right or the left resonance islands, respectively.}, this gives the striking possibility for a particle to pass softly from the left resonance island to the right one, simply by getting \emph{around} the discontinuity line and crossing it in its smooth direction. These trajectories are recognizable on the secular phase portraits as the level curves connected \emph{only} to the \emph{upper side} of the thick green line. Some of them are visible on the graph C of Fig.~\ref{fig:pro-retro1k} (including the trajectory featuring the largest variations of $\tilde{q}$): when crossing the line from the upper part, the single island in which oscillates $\sigma$ becomes the left one (the corresponding secular representation is obtained by $\omega\rightarrow\pi-\omega$). Since the evolution is not constrained by the discontinuity line, that very particular kind of trajectory allows the largest perihelion variations possible for resonances of type $1\!:\!k$.
   
\begin{figure}
      \centering
      \includegraphics[width=\textwidth]{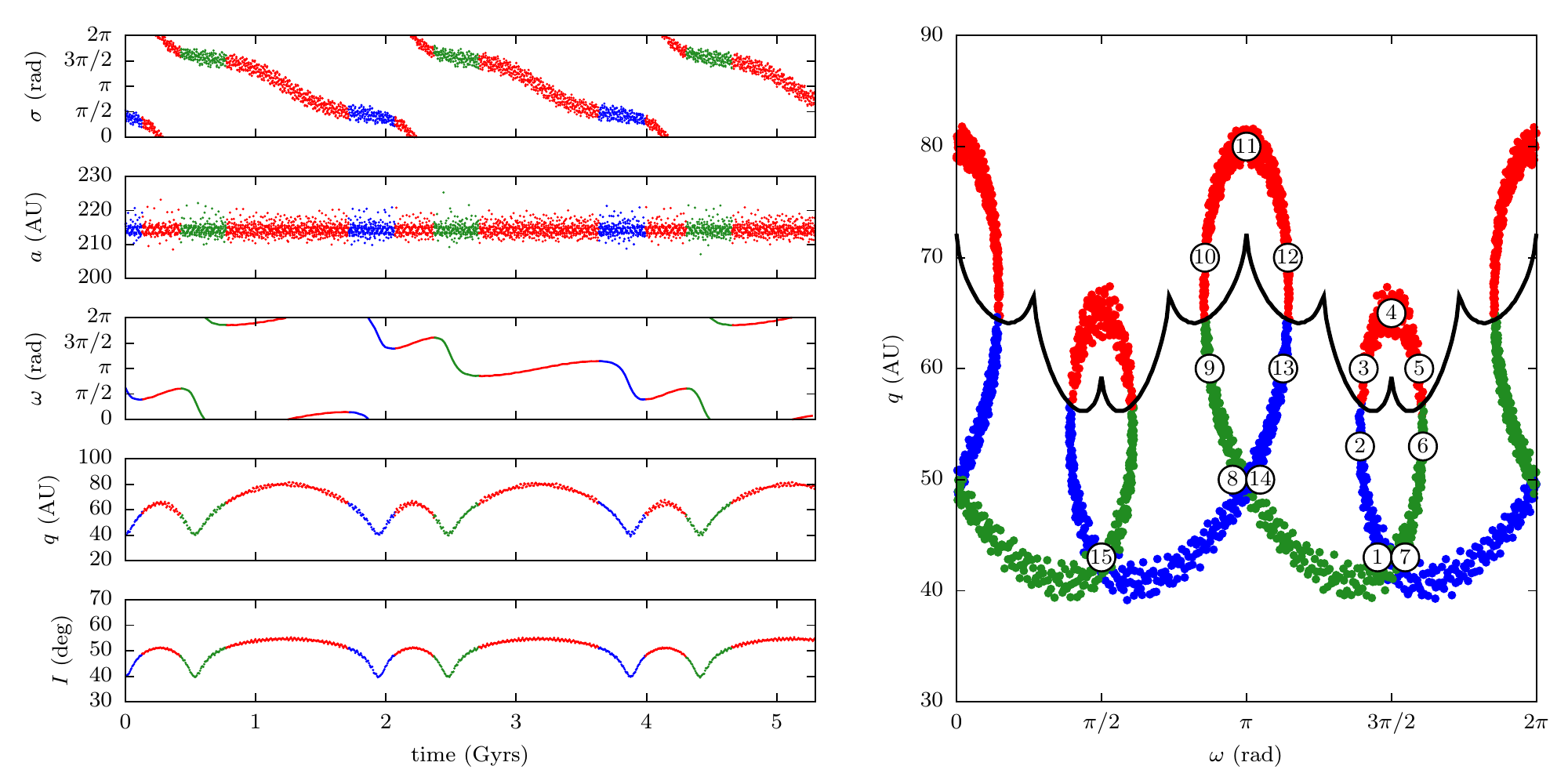}
      \caption{Numerical integration of the unaveraged system for the resonance $1\!:\!19$ with Neptune. On the right, the trajectory is plotted in the plane $(\omega,q)$, where the secular discontinuity line is added in black. The oscillations of the semi-major axis are essentially due to the motion of the Sun around the barycentre of the Solar System: in the averaged system, the secular shift of the resonance centre ranges only from $214.775$~AU to $214.777$~AU. The colour code is chosen according to the geometry of the semi-secular phase space: when it contains two resonance islands, the trajectory is plotted in green (right island) or blue (left island). In the red portions, on the contrary, there is only one resonance island. The numbers refer to Fig.~\ref{fig:nosep}, where the geometry of the semi-secular phase space is shown.}
      \label{fig:osc_perio}
      \vspace{0.5cm}
      \includegraphics[width=0.9\textwidth]{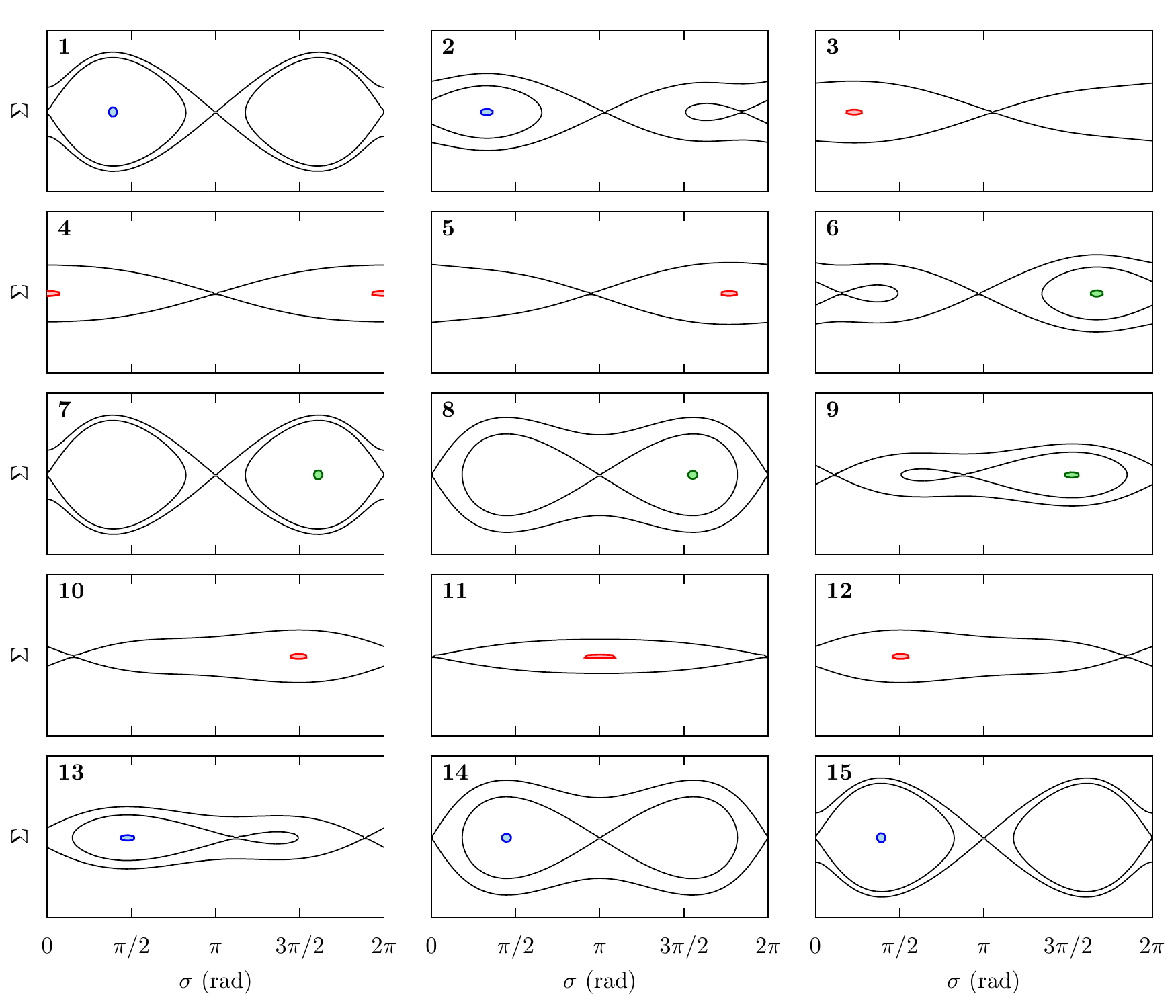}
      \caption{Level curves of the semi-secular Hamiltonian $\mathcal{K}$ for $(U,u)$ fixed according to $15$ points of Fig.~\ref{fig:osc_perio}. For clarity, only the separatrices are shown. We recall that $\Sigma=\sqrt{\mu a}/k$ is the momentum conjugated to $\sigma$. When a separatrix disappears (as between the graphs $2$ and $3$), it means that the smallest resonance island has vanished, so all the level curves enclose now the remaining equilibrium point. The semi-secular trajectory corresponding to the numerical integration of Fig.~\ref{fig:osc_perio} is shown with the same colour code: blue for $\sigma$ oscillating in the left island, green for the right one, and red when there is only one resonance island in the phase space. As the secular parameter $J$ is very close to $0$, the semi-secular trajectory encircles tightly the resonance centre. The tiny area $|2\pi J|$ conserved is shown in mid-tones. Note that between the graphs $1$-$7$ and $8$-$14$, the particle returns exactly to the same $(\omega,q)$ point (thus the same semi-secular geometry) but has changed resonance centre. On a secular time-scale, the corresponding trajectory is a circulation of the resonance centre itself (see the evolution of $\sigma$ on Fig.~\ref{fig:osc_perio}).}
      \label{fig:nosep}
   \end{figure}

   Fig.~\ref{fig:osc_perio} presents an example of such a trajectory obtained by a numerical integration of the unaveraged system (the equations of motion are given by the Hamiltonian \eqref{eq:Hgen} without any transformation). The initial conditions are chosen to match the parameters of Fig.~\ref{fig:Hsec_general_1k} for $\eta_0=0.45$, where that trajectory is noticeable. Naturally, Fig.~\ref{fig:Hsec_general_1k} is plotted for $\sigma$ oscillating in the right resonance island, so only the green and red parts of Fig.~\ref{fig:osc_perio} can be represented. Indeed, for these particular trajectories, a secular representation as described in this article is necessarily piecewise even if the dynamics is perfectly regular. When the secular discontinuity line is crossed, the particle always occupies the non-vanishing island, which prevents any separatrix crossing. That mechanism is detailed on Fig.~\ref{fig:nosep}, where the semi-secular geometry is presented.
   
   \subsection{Higher amplitude oscillations of the resonant angle $\sigma$}\label{subsec:highAmp}
   For a larger area $|2\pi J|$, that is a higher amplitude of the semi-secular oscillations of~$\sigma$, the inevitable separatrix crossing beyond some value of the perihelion distance produces important changes of the resonant secular dynamics (see Paper I). In other words, $\sigma$ is constrained to circulate if the perihelion distance grows too much.
   
\begin{figure}
      \centering
      \includegraphics[width=\textwidth]{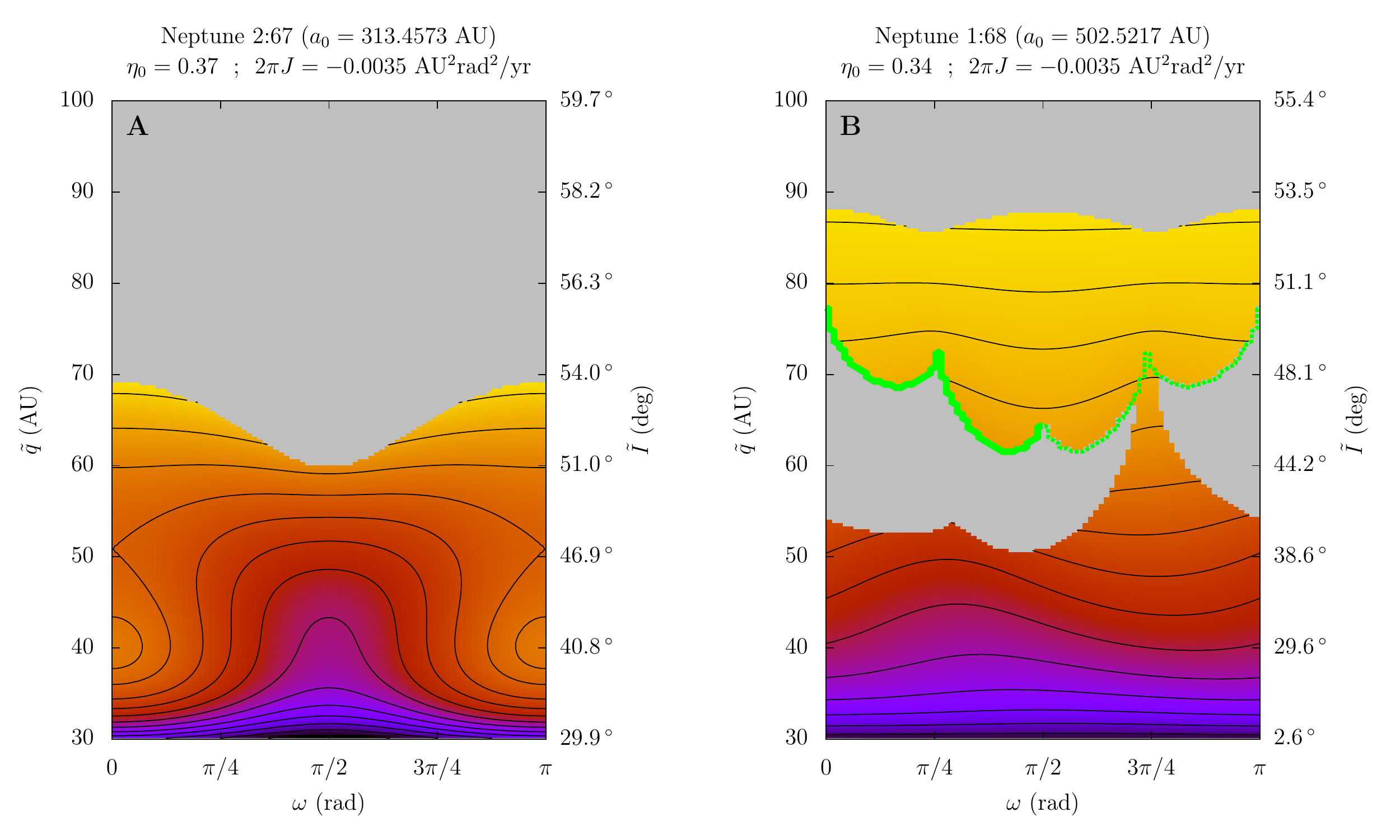}
      \caption{Effect of a large area $|2\pi J|$. The resonances and corresponding $a_0$ are indicated above the graphs, as well as the parameters chosen. The grey colour denotes the regions where the semi-secular separatrices are too narrow to contain the area $|2\pi J|$, that is where that secular model is not relevant ($J$ would have to be redefined). On the graph B ($1\!:\!k$ resonance), the particle is chosen to oscillate inside the right resonance island, and only one island remains beyond the green line. These phase portraits have to be compared to their counterpart with $J=0$ in Fig.~\ref{fig:prograde} and \ref{fig:pro-retro1k} (graphs A and D, respectively).}
      \label{fig:crois}
   \end{figure}

   As shown on Fig.~\ref{fig:crois}, this has the general effect of squeezing the secular trajectories towards Neptune, and even smooth out the equilibrium points if $|2\pi J|$ is big enough. Indeed, a high-amplitude oscillation of $\sigma$ weakens the resonant part of the Hamiltonian function. This lets the non-resonant part dominate, and the latter has no equilibrium point for $a>80$~AU other than the classic Kozai ones around $I=63\degree$ and $117\degree$. In the upper grey regions of the graphs A and B, a secular model for a circulating $\sigma$ can be applied. However, for such models the equilibrium points are rare and concentrated at small values of $\tilde{q}$, where the resonance is still close and effective (see Paper I for an example). Moreover, the circulation of the resonant angle often triggers a diffusion of semi-major axis and the secular representation ceases to be relevant (the locking in resonance acting as a barrier against diffusion). In the lower grey region of the graph B, the particle is pushed out of the right resonance island, but since the left one has still a wide extent, a second capture can happen immediately and produce complex regular-by-parts trajectories. That transition can occur also in the $J=0$ case, but only when the thick green line is crossed, that is at a much higher $\tilde{q}$. A large area $|2\pi J|$ restricts thus severely the possibility for a particle to reach high perihelion distances.
   
   These considerations can be summed up by: deeper the resonance capture, larger the possible secular variations of the perihelion distance. However, the structures at low value of $\tilde{q}$, and in particular the usual equilibrium point at $\omega=0$, can persist even for a large area $|2\pi J|$ since the resonance islands are pretty big for a perihelion near Neptune (see the graph A of Fig.~\ref{fig:crois}).
   
\section{Application to observed objects}\label{sec:app}
   Because of evident observational constraints, the data for high-perihelion transneptunian objects is scarce and subject to large uncertainties. Hence, it is quite hard to determine if a particular observed object in that region presents a mean-motion resonance with one of the planets. Since the dynamics is chaotic in general, even a small change of semi-major axis can lead to very different long-term behaviours, with locking or not in a variety of mean-motion resonances. It is not rare, for example, that an object thought to be in resonance with Neptune for a given best-fit orbit, proved actually to be in a diffusive state when we add new observational constraints. For that reason, we turned to a different approach to the problem: \emph{if} a specific known object \emph{were} in resonance with Neptune (now or in a close past or future), what \emph{would} be its long term dynamics? In particular, would its current orbital elements allow some large variations of its perihelion distance $q$ and/or confine its argument of perihelion $\omega$? Independently of the value of its semi-major axis, its osculating variable $\eta=\sqrt{1-e^2}\cos I$ can indeed be used as parameter for a resonant secular model. We will see that such a general study, which is almost independent of the fit precision of the semi-major axis, can still be very informative.
   
\begin{figure}
      \centering
      \includegraphics[width=\textwidth]{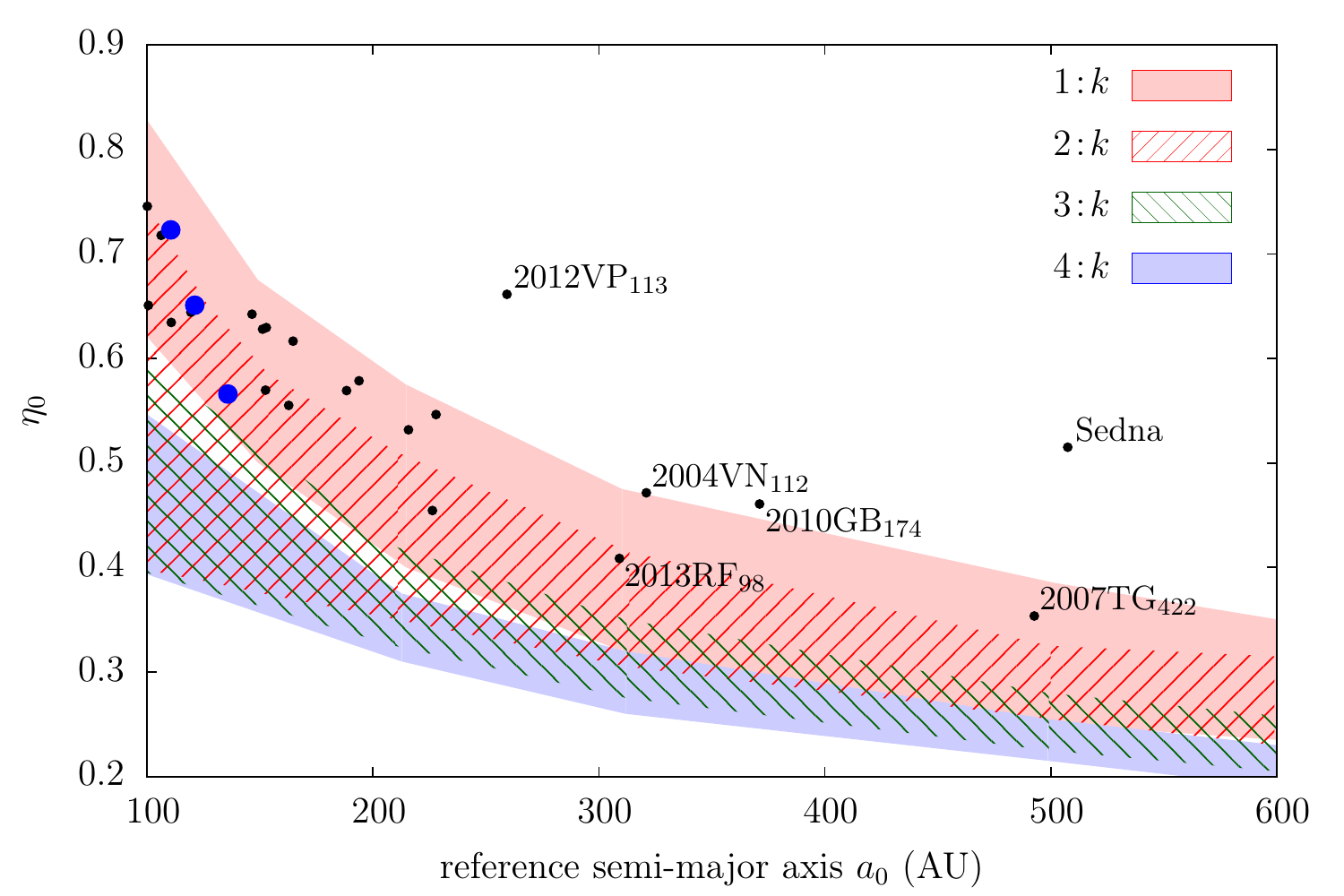}
      \caption{Range of interest (that is the interval of $\eta_0$ allowing libration centres) for various types of mean-motion resonances with Neptune. It is calculated for $J=0$ and the plot is restricted to prograde orbits ($\eta_0>0$). The black spots represent the position of 24 well-known small bodies with $a>100$ AU and $q>30$ AU according to AstDyS database, using $a$ for $a_0$ and $\sqrt{1-e^2}\cos I$ for $\eta_0$. The transneptunian object $2013\text{RF}_{98}$ is added from the JPL database. The names are shown for the bodies with $a>250$ AU, and the blue points refer to those taken in example in the following (from left to right: $303775$, $181902$ and $2007\text{LH}_{38}$).}
      \label{fig:interest}
   \end{figure}

   At first, we can restrict the study to small oscillations of $\sigma$, that is for $J~\approx~0$. As shown in Sect.~\ref{sec:explor}, these trajectories are the most stable and present the most interesting variations of the secular orbital elements. Figure~\ref{fig:interest} presents the ranges of interest of $\eta_0$ for prograde orbits in resonances of types $k_p=1,2,3,4$ with Neptune. They have been obtained by a systematic exploration and a fine tuning of $\eta_0$. As anticipated in Sect.~\ref{subsec:typeGen}, the range of interest follows a definite law with respect to the reference semi-major axis $a_0$ of the resonance (iso-curves of inclination and perihelion distance). The position of the known small bodies with $q>30$ AU and $a>100$ AU are added\footnote{The orbital elements are taken from AstDyS database (hamilton.dm.unipi.it/astdys), except from $2013\text{RF}_{98}$ which comes from the JPL database.}$^,$\footnote{Our secular model can be applied as well to high-perihelion objects with smaller semi-major axes. In particular, it could be used to explore in a plain way the close past and future of the resonant objects recently described by~\cite{SHEPPARD-etal_2016}. This is left for future papers.}, showing in an evident way which ones could have a potentially interesting resonant relation with Neptune. The transneptunian object $2013\text{RF}_{98}$ is also shown despite its very ill-determined orbit because it has been recently used in the article by \cite{BATYGIN-BROWN_2016}. If the semi-major axes of these bodies are not locked in resonance but diffuse slowly, a temporary trapping could also have the effects described by the secular model.
   
\begin{figure}
      \centering
      \includegraphics[width=0.6\textwidth]{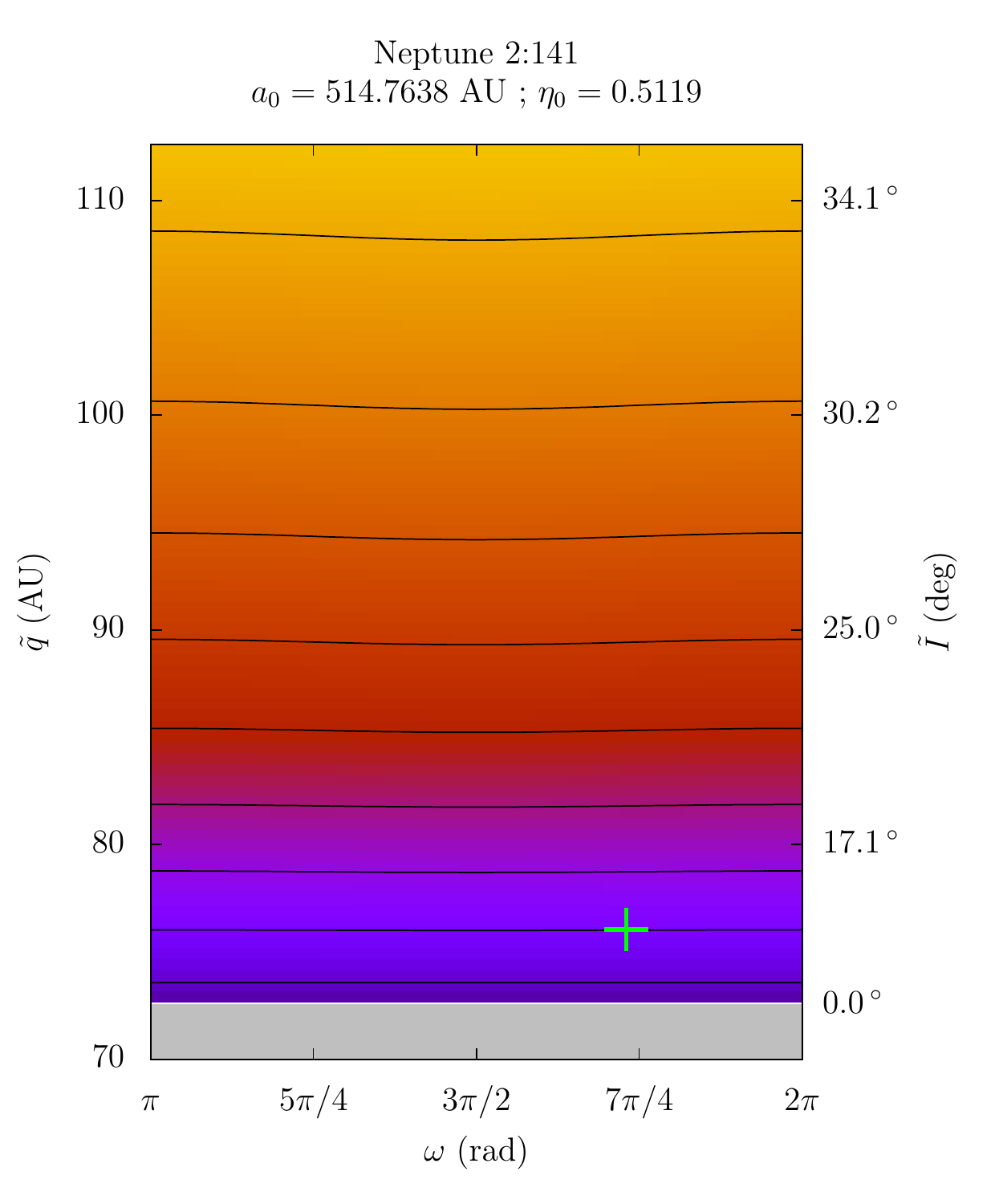}
      \caption{Level curves of the secular Hamiltonian for the resonance $2\!:\!141$ with Neptune, which is located near the semi-major axis of Sedna. The value chosen for $a_0$ is indicated above the figure as well as the parameter $\eta_0$ corresponding to the elements of Sedna. This graph is plotted for $J=0$ and the green cross shows the current position of Sedna in the plane $(\omega,q)$.}
      \label{fig:Sedna}
   \end{figure}

   First of all, we see that Sedna and $2012\text{VP}_{113}$ are completely out of every range of interest (they would require of much higher inclination or a much lower perihelion distance to enter the coloured zones). We can thus affirm that these two bodies \emph{cannot} have had a perihelion near the planetary region and been drifted away by the secular interaction (resonant or non-resonant) with the known planets. A similar result was obtained by numerical means by \cite{GOMES-etal_2005}. As an illustration, Fig.~\ref{fig:Sedna} shows the secular behaviour that Sedna would have if it were in a resonance of type $2\!:\!k$ with Neptune with $J=0$. Naturally, that graph corresponds to a specific resonance, but Figure~\ref{fig:interest} certifies that \emph{every other} resonance of the neighbourhood would produce such flat level curves. Moreover, numerical integrations show that for low inclinations and perihelion distances near Neptune, the resonances with $a_0 > 350$~AU are unstable: if an object happens to stay in such resonances, its perihelion distance should be \emph{already} quite high to avoid any diffusion of semi-major axis\footnote{In particular, we did not manage to lock $2007\text{TG}_{422}$ in resonance in our numerical integrations (semi-major axis $\approx 493$ AU) even by putting it exactly at the resonance centre. This is different for high inclinations, for which it is not rare to observe stable resonance captures with $a_0 > 500$~AU even for perihelion distances near Neptune. However, the probability to find a real body with that kind of orbit is likely very low.}. Eventually, one can think of a more complex scenario for Sedna and $2012\text{VP}_{113}$, including an initial resonant semi-major axis of $150$-$200$ AU (see Fig.~\ref{fig:interest}), followed by an excursion of perihelion, and finally a diffusion of $a$. However, this is also impossible because the semi-major axis cannot be affected by diffusion once the perihelion is so high, even if the particle is pushed out of the resonance \citep[remember the empirical law of $q_\text{min} = a/27.3 + 33.3$~AU by][]{GALLARDO-etal_2012}. Such a mechanism is definitely not able to explain the current orbits of these two transneptunian objects. We should think of other scenarios (as the interaction with an unknown external perturber) or different sources, as the Oort Cloud or a neighbour star in the birth cluster of the Sun \citep[see in particular][and their other related works]{JILKOVA-etal_2015}.
   
   On the other hand, many objects are located well inside some range of interest, and sometimes even simultaneously for several types of resonance. In the following, we show various secular phase portraits that some of these small bodies would have if they were in resonance with Neptune. We chose only the transneptunian objects compatible with a deep resonance, that is those for which the current value of the mean anomaly leads to a resonant angle $\sigma$ near the equilibrium. We can get an estimation of the minimum area $|2\pi J|$ an object could have by assuming that its current secular semi-major axis is $a_0$ and considering the corresponding surface in the $(\Sigma,\sigma)$ plane. This was done for specific resonances, but every neighbour resonance of the same type would produce a similar phase portrait (see Sect.~\ref{sec:explor}). Each graph presents also the result of a numerical integration including the four giant planets. The secular variations of their orbital elements are given by the synthetic representation of~\cite{LASKAR_1990}, supposed valid in the entire duration of the integrations. The initial conditions are taken from the best-fit orbits of AstDyS, excepted for the semi-major axes which are adjusted to produce the resonant captures. The required modifications are of the order of $2$ AU, which is generally larger than the $3$-$\sigma$ uncertainty given by AstDyS. Hence, the trajectories shown are not supposed to represent the ``real" motion of these objects, but only to give an insight of what could be their secular dynamics in case of resonance.
   
\begin{figure}
      \centering
      \includegraphics[width=\textwidth]{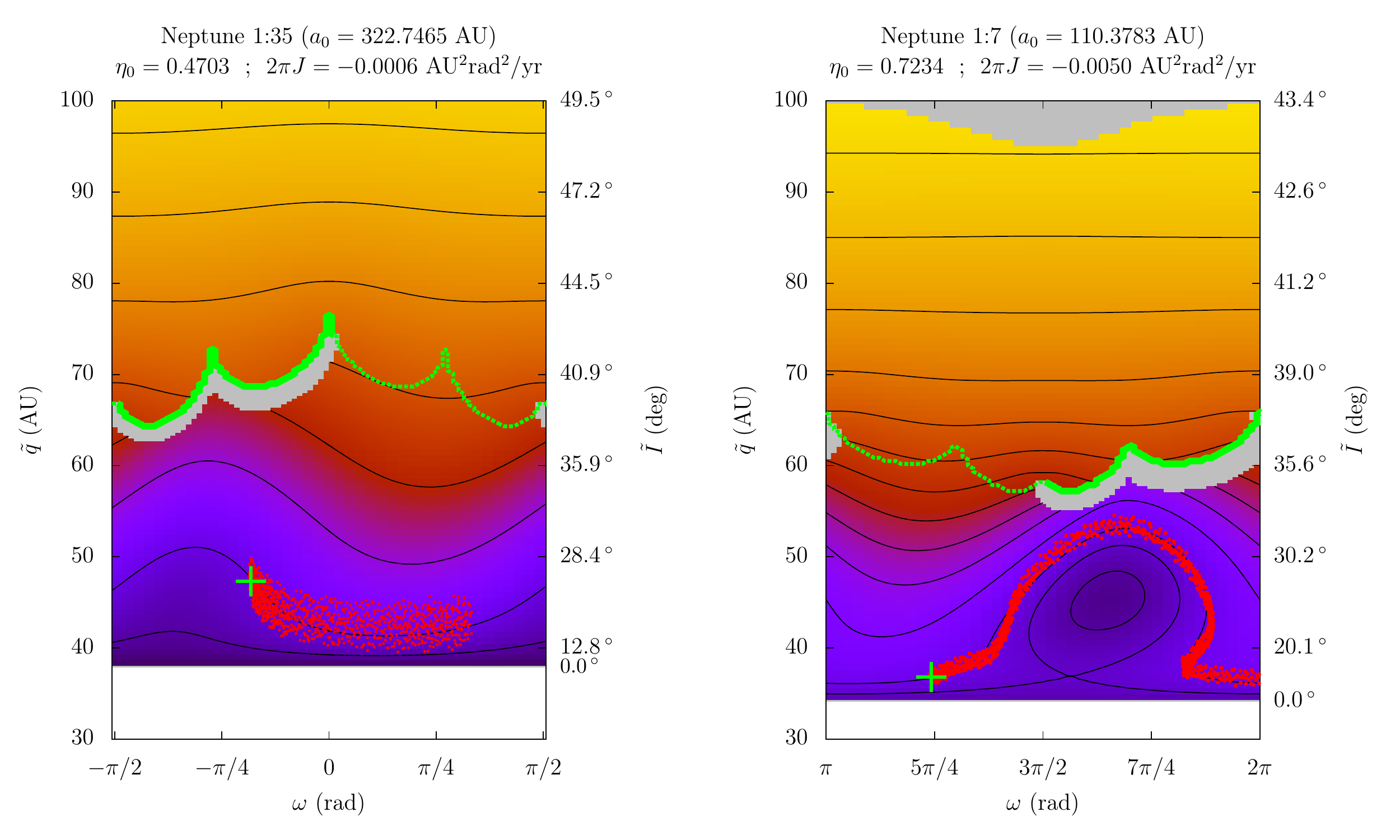}
      \caption{Level curves of the secular Hamiltonian for two mean-motion resonances with Neptune located near the semi-major axes of the transneptunian objects $2004\text{VN}_{112}$ (left) and $303775$ (right). The chosen values for $a_0$ are indicated above the figures as well as the parameters corresponding to the elements of these bodies. The green crosses show their current positions in the plane $(\omega,q)$ and the red dots are the results of numerical integrations: the presented time-spans are $350$ Myrs on the left and $200$ Myrs on the right. The secular evolutions on a larger time-scale are found to be rather periodic.}
      \label{fig:2004VN112-303775}
   \end{figure}

   The left graph of Fig.~\ref{fig:2004VN112-303775} presents the case of $2004\text{VN}_{112}$ which is located at the limit of the range of interest for $1\!:\!k$ resonances. Indeed, no equilibrium point remains and $\omega$ circulates. This kind of resonance would only result in oscillations of the perihelion of $2004\text{VN}_{112}$ from about $42$ to $50$ AU. The right graph of Fig.~\ref{fig:2004VN112-303775} shows the case of $303775$ which is well inside the range of interest for $1\!:\!k$ resonances. Its current position would imply a reachable perihelion distance of $52$ AU. Note that $303775$ is also in the zone of interest for $2\!:\!k$ resonances, but its current mean anomaly would imply a large oscillation of the resonant angle leading to an unstable capture.
   
\begin{figure}
      \centering
      \includegraphics[width=\textwidth]{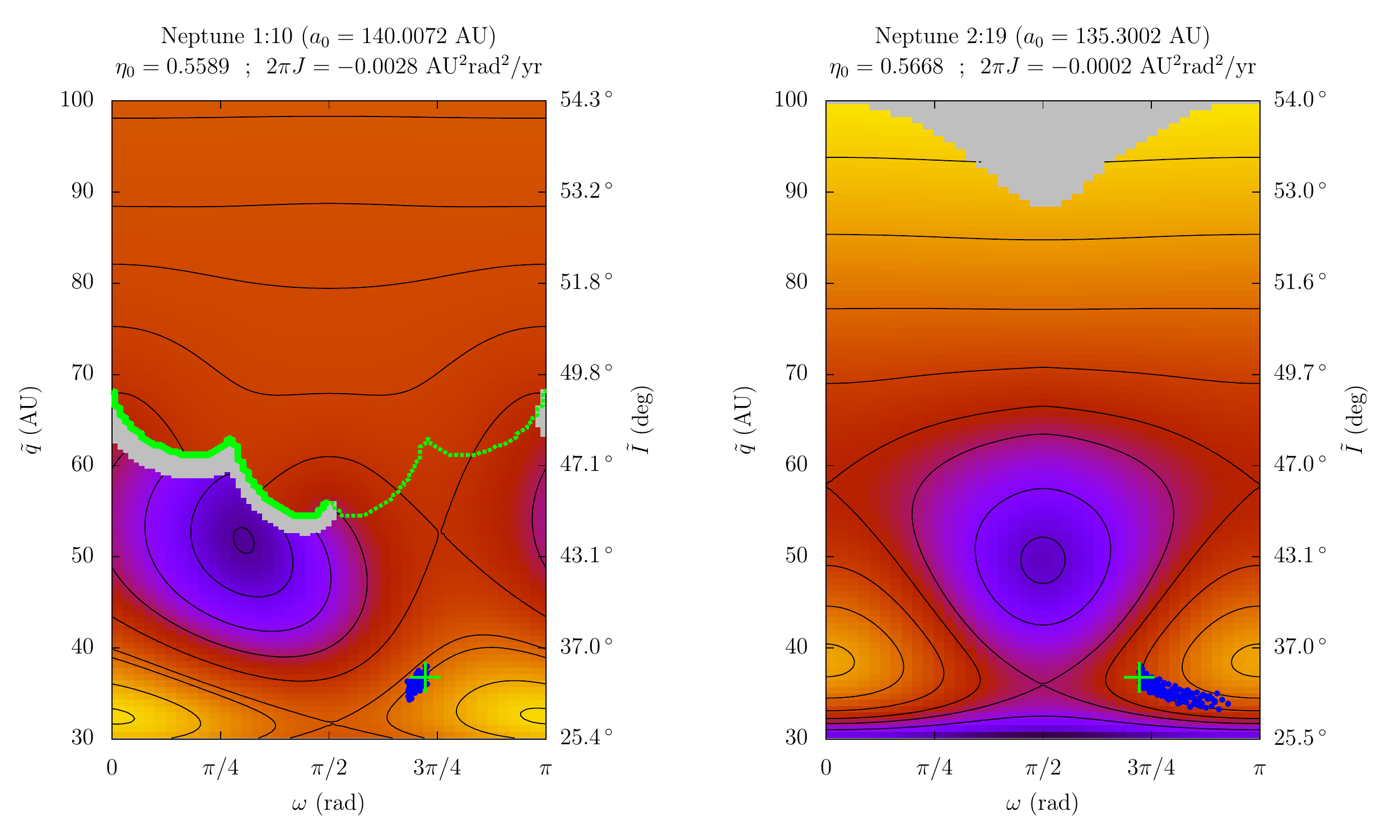}
      \caption{Level curves of the secular Hamiltonian for two mean-motion resonances with Neptune located near the semi-major axis of the transneptunian object $2007\text{LH}_{38}$. The chosen values for $a_0$ are indicated above the figures as well as the parameters corresponding to the elements of $2007\text{LH}_{38}$. The green crosses show its current position in the plane $(\omega,q)$ and the blue dots are the results of numerical integrations: the presented time-spans are $14$ Myrs on the left and $28$ Myrs on the right, after which $2007\text{LH}_{38}$ is pushed outside of the resonances considered.}
      \label{fig:2007LH38}
   \end{figure}

   On Fig.~\ref{fig:2007LH38}, we see that resonances of types $1\!:\!k$ and $2\!:\!k$ would both decrease the perihelion distance of $2007\text{LH}_{38}$ towards Neptune, where the overlap with neighbour resonances leads eventually to a chaotic diffusion of the semi-major axis (not shown here). A very similar behaviour is observed for resonances near the nominal orbit of $2013\text{RF}_{98}$, but its very large uncertainties make very wide the range of possible resonances involved.
   
\begin{figure}
      \centering
      \includegraphics[width=\textwidth]{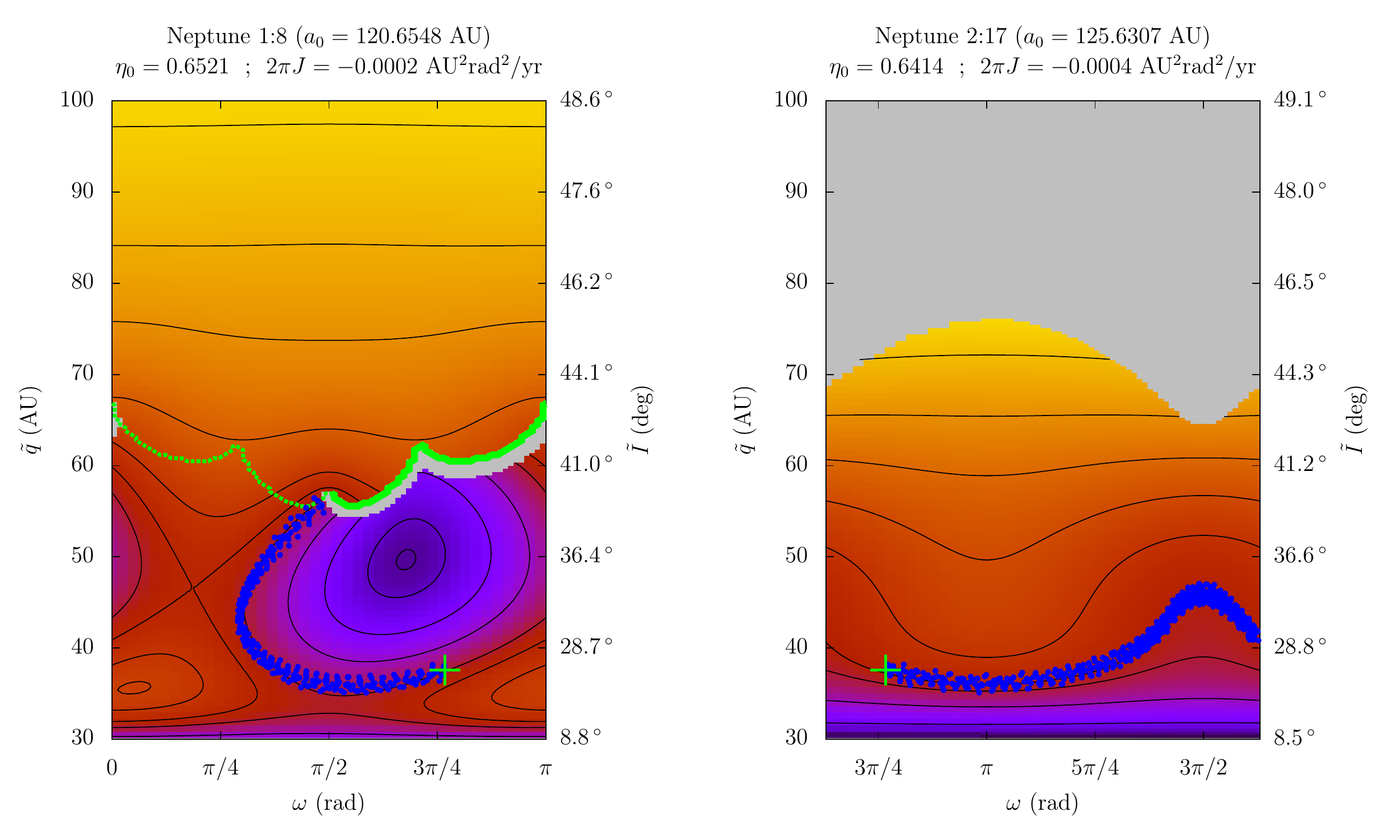}
      \caption{Level curves of the secular Hamiltonian for two mean-motion resonances with Neptune located near the semi-major axis of the transneptunian object $181902$. The chosen values for $a_0$ are indicated above the figures as well as the parameters corresponding to the elements of $181902$. The green crosses show its current position in the plane $(\omega,q)$ and the blue dots are the results of numerical integrations: the presented time-spans are $100$ Myrs on the left and $30$ Myrs on the right. On the left graph, the secular evolution leads the particle to the discontinuity line, where the occupied resonance island disappears. After the transition, the particle goes on switching secular behaviour (not shown here) before finally being trapped at $q\approx 62$~AU. On the right graph on the contrary, the secular evolution on a larger time-span is found to be rather periodic.}
      \label{fig:181902}
   \end{figure}

   Finally, Figure~\ref{fig:181902} shows that a resonance of type $2\!:\!k$ would result in small-amplitude oscillations of the perihelion of $181902$ along with a circulation of $\omega$. On the contrary, a resonance of type $1\!:\!k$ would raise its perihelion distance as far as $56$ AU and possibly maintain it high on a billion-year time-scale, depending on the next resonant behaviours adopted by the particle (see Paper I). Section~\ref{sec:trapping} shows that this is a common mechanism to produce long-lived small bodies with high perihelion distances. Even if $181902$ is not considered to follow that kind of dynamics, Fig.~\ref{fig:181902} shows that it is located in the required range of orbital elements, demonstrating that such a behaviour is not only a mathematical curiosity of the resonant secular model.
   
\section{High-perihelion trapping mechanisms}\label{sec:trapping}   
\begin{figure}
      \centering
      \includegraphics[width=\textwidth]{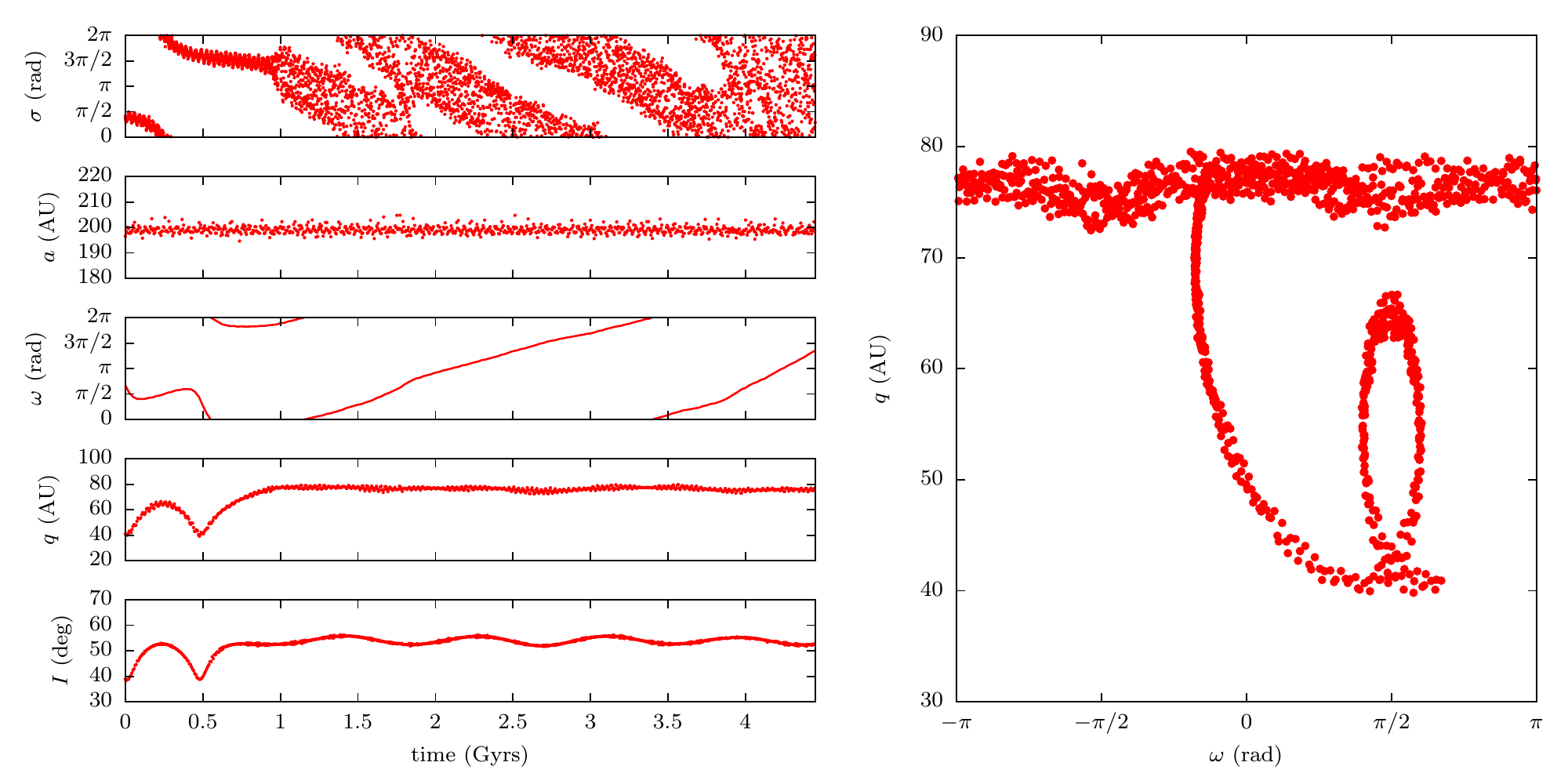}
      \caption{Example of a particle initially trapped in the $1\!:\!17$ mean-motion resonance with Neptune. This numerical integration includes the four giant planets with the secular variations of their orbital elements. On the right, the same trajectory is plotted in the $(\omega,q)$ plane. It should be compared with Fig.~\ref{fig:osc_perio} (neighbouring resonance and similar initial conditions), for which the circular and coplanar planetary orbits make the resonant dynamics much more stable, even at high perihelion distances.}
      \label{fig:cutoff}
   \end{figure}

   The possibility of transferring Scattered Disc objects to high perihelion distances by resonant interactions with Neptune is well known since the work of \cite{GOMES-etal_2005}. Our resonant secular model is a suitable tool to further precise their results. In particular, they mention a possible trapping caused by the drop-off of the resonant terms in the disturbing function when the perihelion distance increases. Indeed, the particle becomes vulnerable to any other kind of perturbation, which destabilizes the resonant dynamics. We give an example of such mechanism on Fig.~\ref{fig:cutoff}, where we chose on purpose a trajectory avoiding the discontinuity line (see Sect.~\ref{subsec:playDisc}). The secular dynamics is perfectly regular during the first loop, because the perihelion excursion is relatively modest. Then, when the resonant link with Neptune gets weaker, the perturbations induced by the varying small inclinations and eccentricities of the planets are strong enough to break the smooth resonant trajectory (at $t\approx 1$ Gyr): the secular parameter $J$ makes unpredictable jumps and the particle is eventually pushed out of the resonance. Since the perihelion distance is very high, the semi-major axis is not subject to diffusion and so the particle remains trapped. The resonance is still very close, however, and the resonant angle switches chaotically from high-amplitude oscillations to circulation, but this has no notable effect on the orbital elements. Indeed, the probability to recover a secular trajectory leading to small values of $q$ is extremely low (it would require a parameter $J\approx 0$).
   
   Actually, that scenario was found to be relatively rare in our numerical experiments because of the very high excursion of $q$ required. However, we found another trapping mechanism coming directly from our conclusions of Sect.~\ref{sec:explor}. In that second scenario, the capture is not due to a drop-off of the resonant terms but simply to the crossing of the secular discontinuity line common to all resonances of type $1\!:\!k$. After the discontinuity, the particle can in fact remain in resonance but on a periodic trajectory avoiding any further crossing of the line in its discontinuous direction (as we described in Sect.~\ref{subsec:playDisc}). In other words, the initial transition triggers an ``irreversible" smooth behaviour, for which the return to the entrance configuration does not imply a new separatrix crossing. It will still be separatrix-grazing, but the natural evolution of the semi-secular phase space will immediately lead the particle away again. The trajectory adopted has a very wide area $|2\pi J|$, because the particle has been pushed outside of the vanishing island while the other one had a wide extent. Then, it simply becomes a horseshoe orbit when the second island reappears, avoiding any further separatrix crossing (the growing island appears \emph{inside} the trajectory). As an illustration, one can imagine on Fig.~\ref{fig:nosep} a trajectory remaining always inside the outer separatrix, but outside the inner one when it appears. Precisely, we saw in Sect.~\ref{subsec:highAmp} that for a large parameter $|2\pi J|$, the secular level curves are very flat, without any large variations of the perihelion distance: this means that the particle reaches a permanent smooth high-perihelion evolution\footnote{Naturally, as the final trajectory relies on a high-amplitude oscillation of the resonant angle, the proximity of the separatrices can lead to an accidental extra transition. However, this proves to be very rare and only temporary, as seen in the next section.}. As shown by numerical simulations, that mechanism is rather frequent for the resonant particles attaining the discontinuity line. Naturally, it cannot involve resonances of type $2\!:\!k$ and further, because the corresponding secular trajectories are all periodic (there is no discontinuity line, see Sect.~\ref{subsec:typeGen}). To get specific examples, one can anticipate a bit and look at Figs.~\ref{fig:integMarc1}-\ref{fig:integMarc3}.
   
   These two mechanisms imply that even a set of non-migrating planets can produce a permanent high-perihelion reservoir, continuously supplied with new objects which have a very low probability to come back to smaller perihelion distances during the lifetime of the Solar System. These objects are added to the primordial population of high-perihelion bodies, left on non-resonant orbits by the migration of Neptune \citep{GOMES-etal_2005}.
   
\begin{figure}
      \centering
      \includegraphics[width=\textwidth]{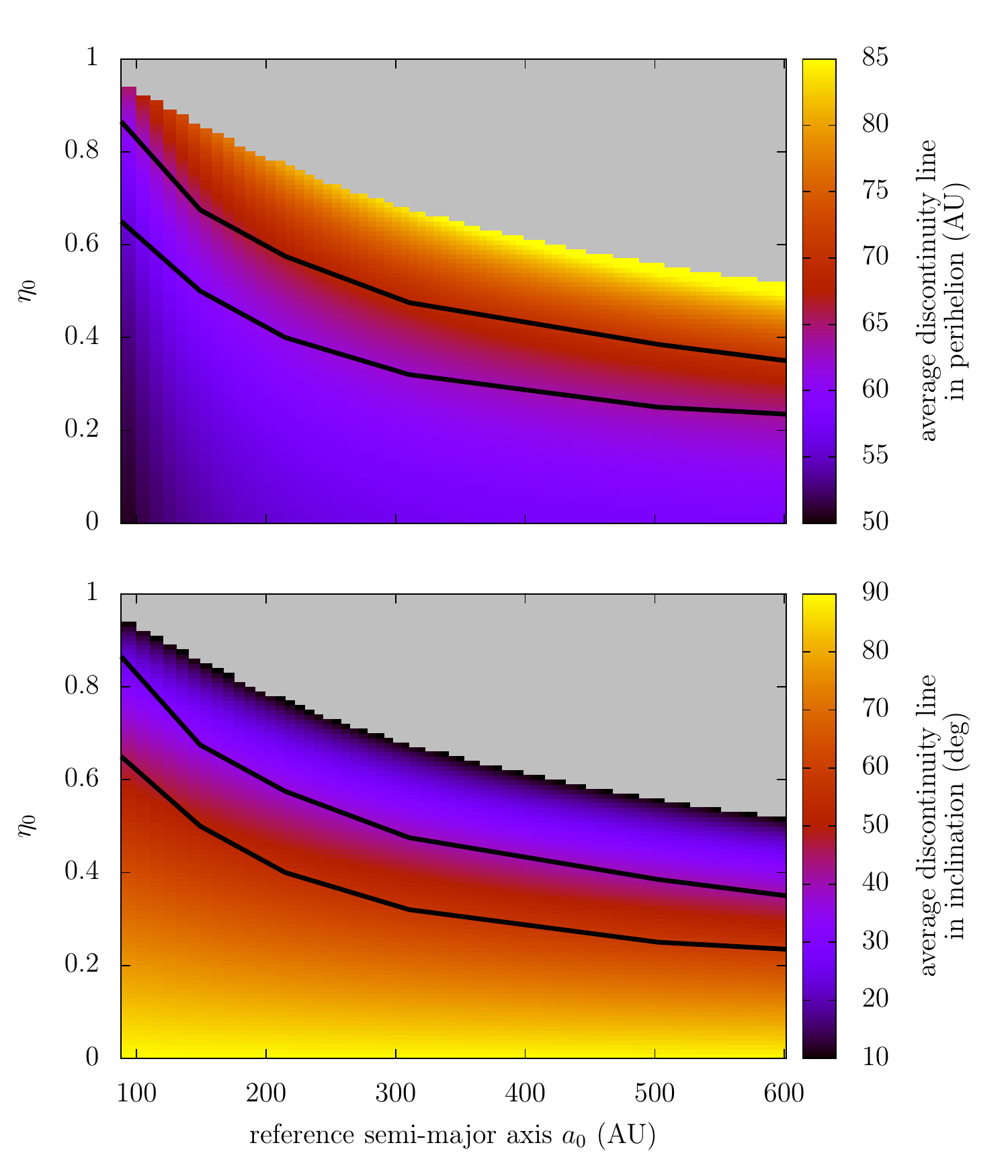}
      \caption{Average position of the secular discontinuity line for all resonances of type $1\!:\!k$ with Neptune from $1\!:\!5$ to $1\!:\!89$. In the grey region, there is no discontinuity line in the secular plots. The black lines delimit the ``range of interest", that is the interval of $\eta_0$ allowing libration centres. A high-perihelion trapping is only possible between these lines, where secular level curves can lead the particle from low perihelion values to the discontinuity line.}
      \label{fig:HPtrapping}
   \end{figure}

   Our secular model can be used to estimate the size and location of the high-perihelion reservoir produced by the second mechanism. For this purpose, we plotted the secular discontinuity line of all the resonances of type $1\!:\!k$ from $a_0=90$ to $600$ AU in a grid of parameters $\eta_0$ in $[0;1]$ (prograde orbits). Since the crossings can happen at all values of $\omega$, we retained only the average line, judged to be representative of a typical perihelion value for the capture (for example $\tilde{q}\approx 60$~AU and $\tilde{I}\approx 41\degree$ for the left graph of Fig.~\ref{fig:181902}). The result is shown on Fig.~\ref{fig:HPtrapping}, both for perihelion distance and inclination. We added the limits of the range of interest for $1\!:\!k$ resonances (same as Fig.~\ref{fig:interest}), because such a trapping can occur only if there are secular level curves \emph{leading to} the discontinuity line. Naturally, a separatrix crossing can happen much before the discontinuity line (see the graph B of Fig.~\ref{fig:crois}), but at low perihelion distances the resonant dynamics is unstable for large values of $|2\pi J|$ and a separatrix crossing often triggers a diffusion of semi-major axis. Hence, in practice the high-perihelion trappings occur indeed near the secular discontinuity line. The reservoir is rather well delimited in the perihelion-inclination space because the position of the average discontinuity line follows more or less the range of interest: from Fig.~\ref{fig:HPtrapping}, we can estimate its extension as roughly $q\in [55;70]$~AU and $I\in [30;50]\degree$. This is consistent with the results of \cite{GOMES-etal_2005}, although their approach was rather different (they counted the numbers of objects ending up at $q>40$~AU in their numerical simulations, whatever the mechanism that led them there). Naturally, these limits do not mean that an object cannot reach higher perihelion distances (see for instance the graph C of Fig.~\ref{fig:pro-retro1k}), but simply that an accumulation of objects should be observed there. Finally, as the perihelion-raising phases of these trajectories are pretty short compared to their initial or final states, a lack of objects in resonance $1\!:\!k$ with Neptune should be observed in the intermediate region (say from $q=40$ to $55$~AU).
   
\section{Incoming objects from the Oort Cloud}\label{sec:OC}
   The Oort Cloud is a well-known source of ``new comets" and more generally of small bodies arriving in the planetary region on very eccentric orbits. There is an extensive literature about the flux of comets traversing the observable zone (near the orbit of the Earth) and the different ways to cross the Jupiter-Saturn barrier, but very little about the contribution of the Oort Cloud to the Scattered Disc. Actually, the combined effects of the planetary perturbations and the galactic tides are an efficient mechanism to continuously replenish the Scattered Disc, and in particular, contribute to the accumulation zone described in the previous section. The main effect of the galactic tides is a long-period oscillation of the perihelion distance, whereas the planets produce the well-known diffusion of semi-major axis. As the galactic tides are only effective for $a>1000$ AU, a typical scenario to create a Scattered Disc object from the Oort Cloud is to drive the perihelion distance a little beyond Neptune, where the planetary perturbations make the semi-major axis decrease below $1000$ AU, turning off the action of the galactic tides. The particle becomes part of the Scattered Disc, where it can be possibly captured in a mean-motion resonance of type $1\!:\!k$ with Neptune and eventually end up in the accumulation zone.
   
\begin{figure}
      \centering
      \includegraphics[width=\textwidth]{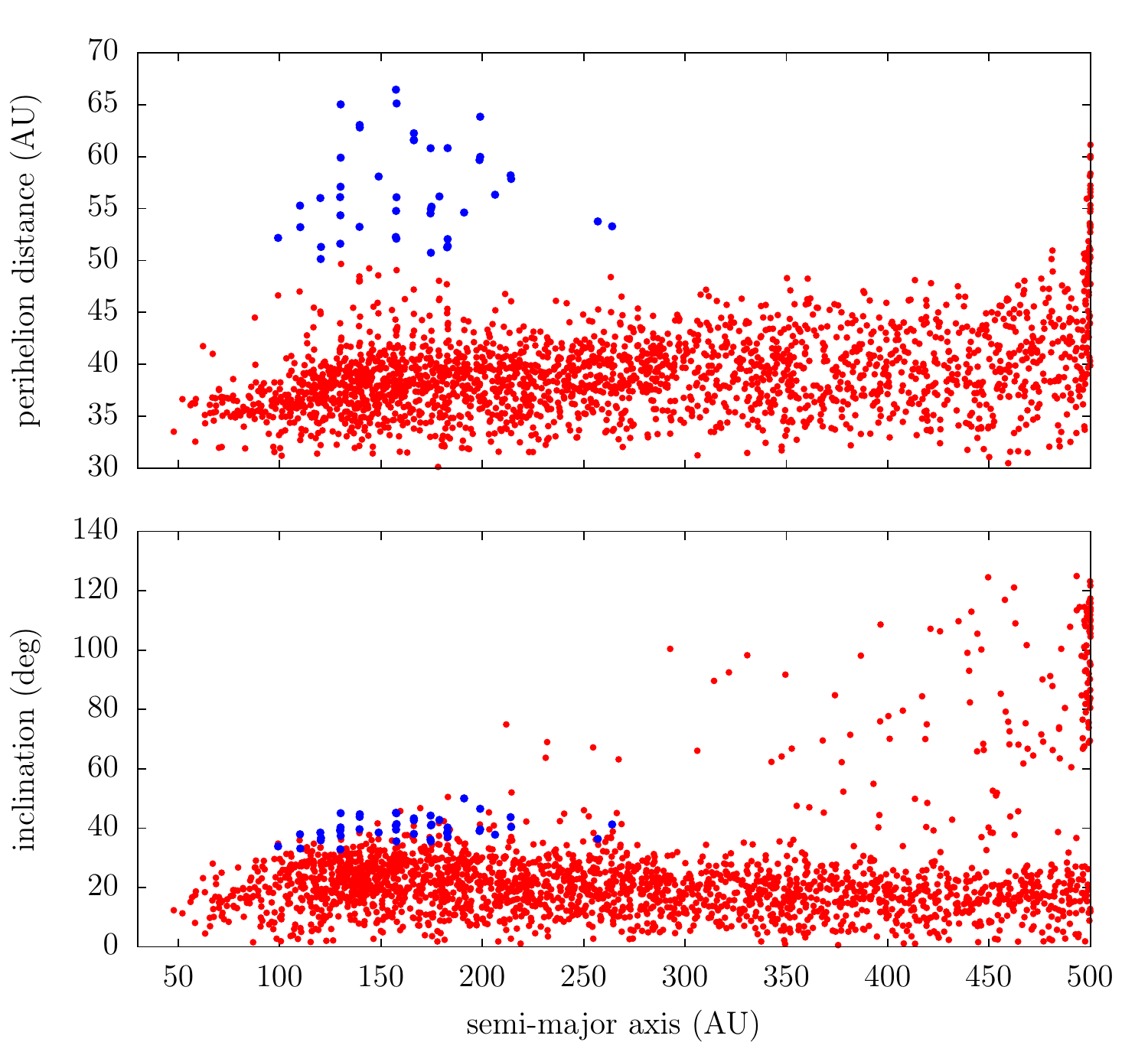}
      \caption{Final state of the sample of particles transferred from the Oort Cloud to the Scattered Disc ($q>30$~AU and $a<500$~AU). The blue points are particles with perihelion distances larger than $50$~AU and semi-major axes smaller than $450$~AU. All of them are locked in mean-motion resonance with Neptune, including only one particle in a resonance of type $2\!:\!k$. The other ones are all locked in resonances of type $1\!:\!k$ and reached their current perihelion distances and inclinations billion years ago thanks to the trapping mechanism (the first one arrived $3$~Gyrs ago and did not move ever since). For semi-major axes larger than $450$~AU, the high-perihelion particles denote the beginning of the Oort Cloud \citep[see][for a wider scale]{FOUCHARD-etal_2016}.}
      \label{fig:reservoir}
   \end{figure}

   The reservoir is actually pretty visible in the results by \cite{FOUCHARD-etal_2016} (see their Fig.~4), although they do not describe it in details. They simulated a precursor of Oort Cloud consisting in $10^7$ objects with initial orbital elements such that $a\in[1100;50\,000]$~AU, $q\in[15;32]$~AU and $I\in[0;20]\degree$. In order to trace the high-perihelion trapping mechanism in their simulation, we picked up the objects arriving into the Scattered Disc (when their semi-major axes become smaller than $500$ AU), and used it as initial conditions for a numerical integration until the date J2000. The corresponding initial times range from the formation of the Solar System until today. We used a more realistic planetary model than in \cite{FOUCHARD-etal_2016}, including the four giant planets with the secular variations of their orbital elements \citep[synthetic representation of][supposed valid in the entire duration of the integration]{LASKAR_1990}. Our results are plotted on Fig.~\ref{fig:reservoir}, centred in the region of interest in the scope of this paper. At first, we see that we should redefine the lower limit of the accumulation zone at $q\sim 50$~AU (instead of $55$) because a significant amount of objects reach the discontinuity line below its average. Naturally, the reservoir is also delimited in semi-major axis, because extremely high-order resonances are unstable and associated with time-scales much longer than the age of the Solar System. According to Fig.~\ref{fig:reservoir}, the mean-motion resonances are efficient to drive the objects into the accumulation zone for $a\in[100;300]$ AU.
   
\begin{figure}
      \centering
      \includegraphics[width=\textwidth]{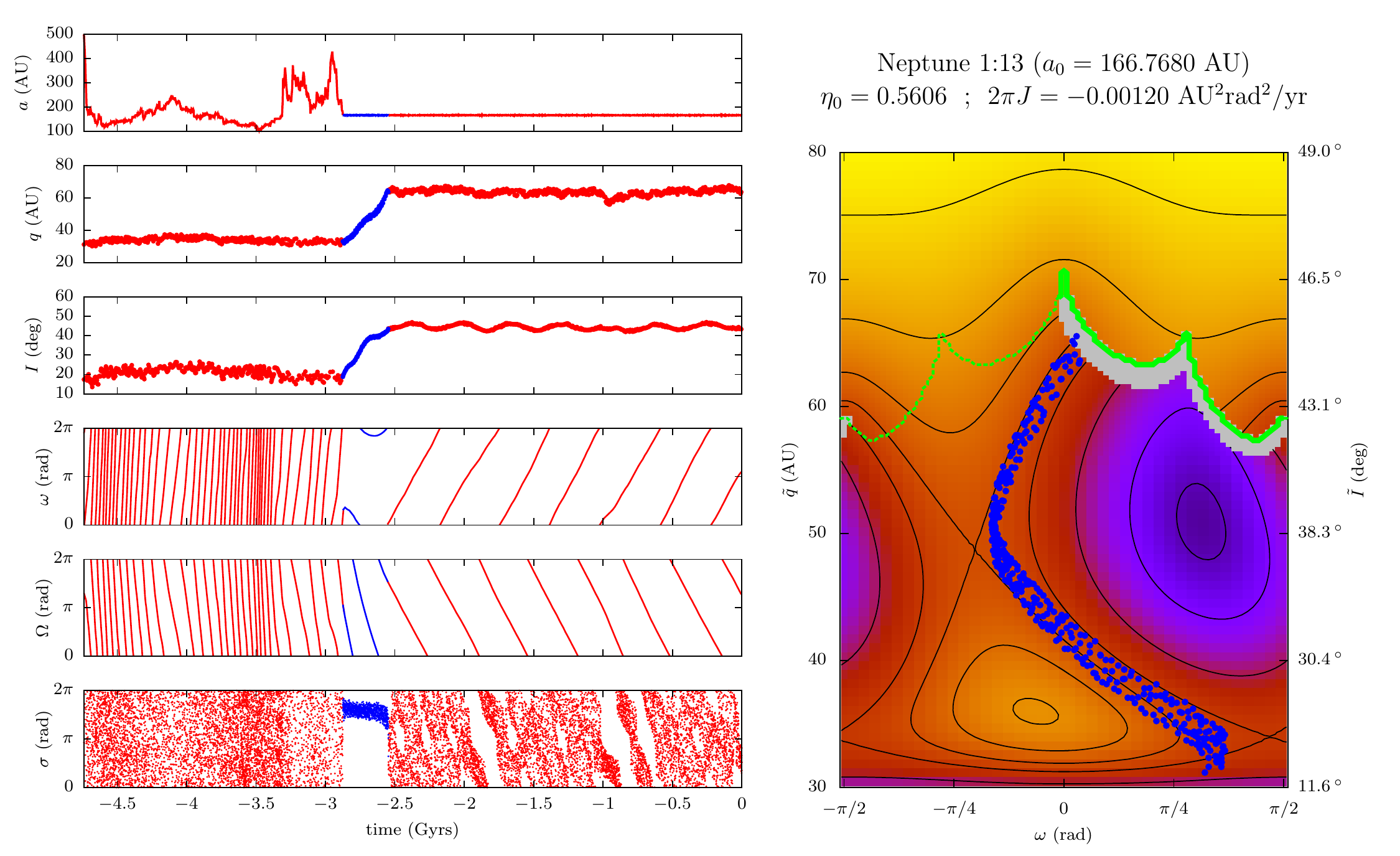}
      \caption{Example of a particle coming from the Oort Cloud and captured in a resonance of type $1\!:\!k$ with Neptune. The origin of time is the date J2000. The resonant dynamics brings the particle toward the secular discontinuity line where it is trapped into the high-perihelion reservoir. On the right, the blue part of the trajectory is shown in the plane $(\omega,q)$ in front of the level curves given by the resonant secular model (the parameters used are indicated above the graph).}
      \label{fig:integMarc1}
   \end{figure}

\begin{figure}
      \centering
      \includegraphics[width=\textwidth]{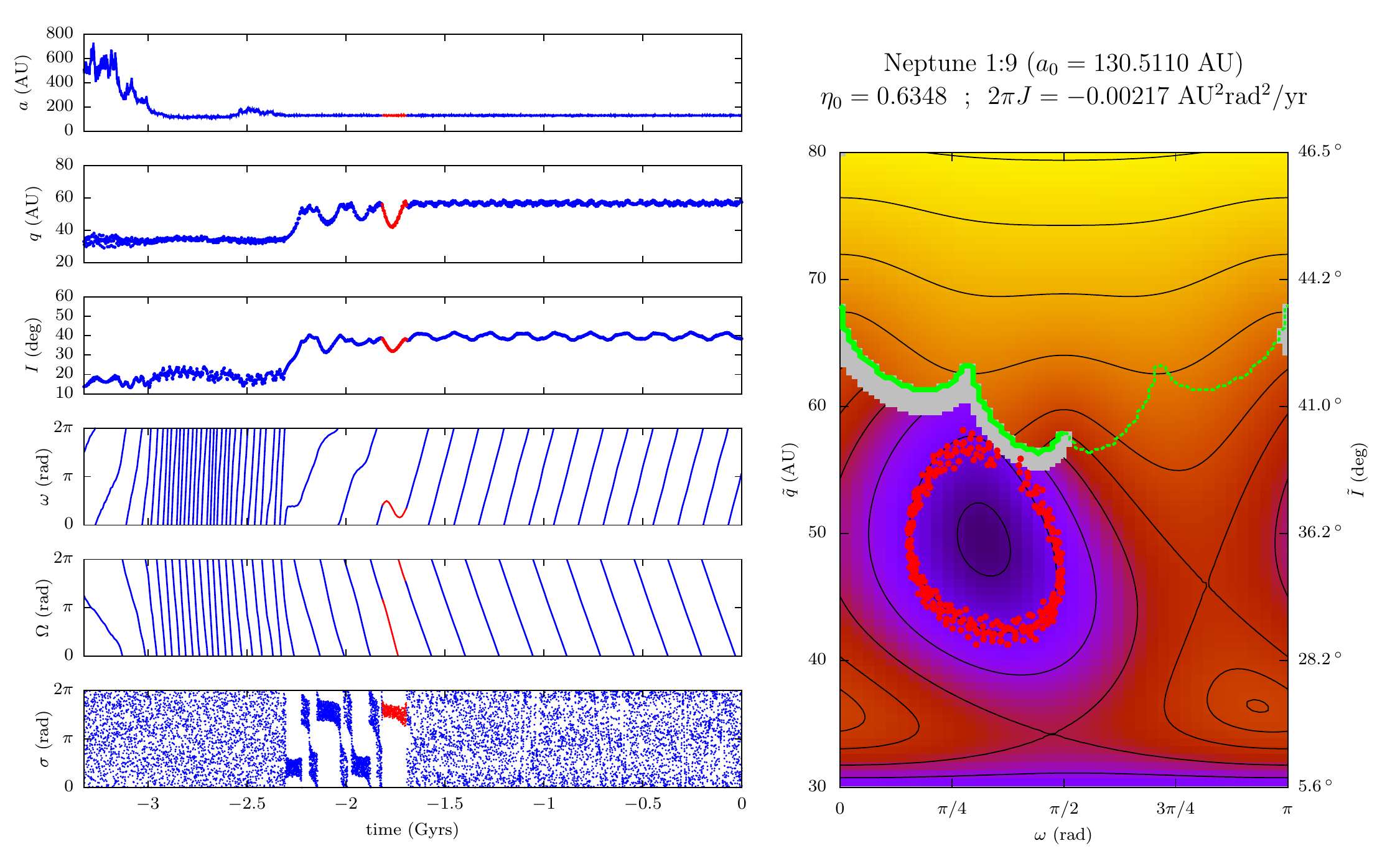}
      \caption{Same as Fig.~\ref{fig:integMarc1} for a particle which is not trapped immediately after the rise of its perihelion distance. Some extra separatrix crossings occur before reaching the long-term stability state specific to the accumulation zone. After the trapping, $\sigma$ oscillates with a high amplitude around a circulating centre (even if the points seem to cover the entire range from $0$ to $2\pi$).}
      \label{fig:integMarc2}
   \end{figure}

\begin{figure}
      \centering
      \includegraphics[width=\textwidth]{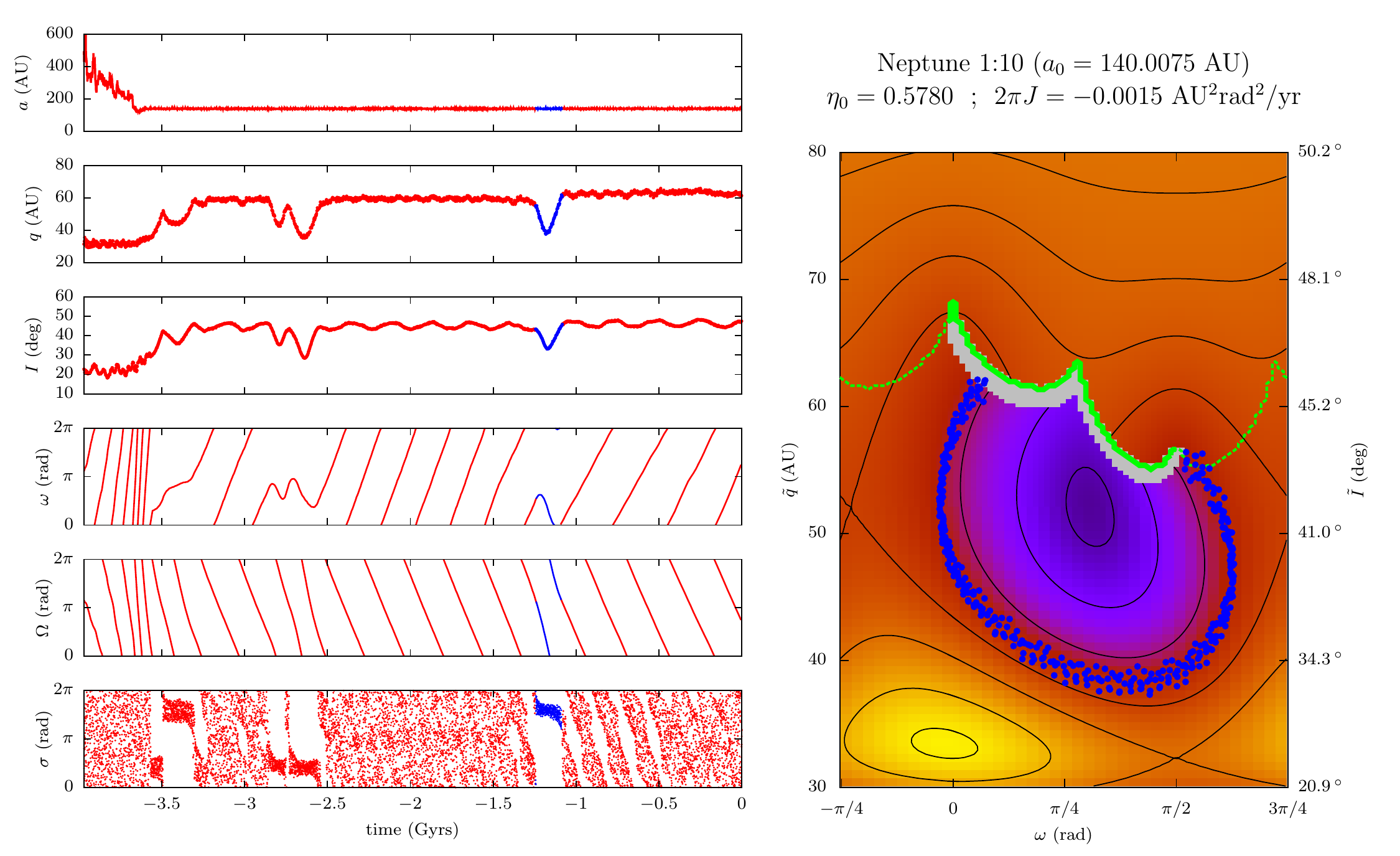}
      \caption{Same as Fig.~\ref{fig:integMarc1} for a particle incurring temporary escapes out of the reservoir. The time spent in the accumulation zone is though much longer than these random excursions.}
      \label{fig:integMarc3}
   \end{figure}

\begin{figure}
      \centering
      \includegraphics[width=\textwidth]{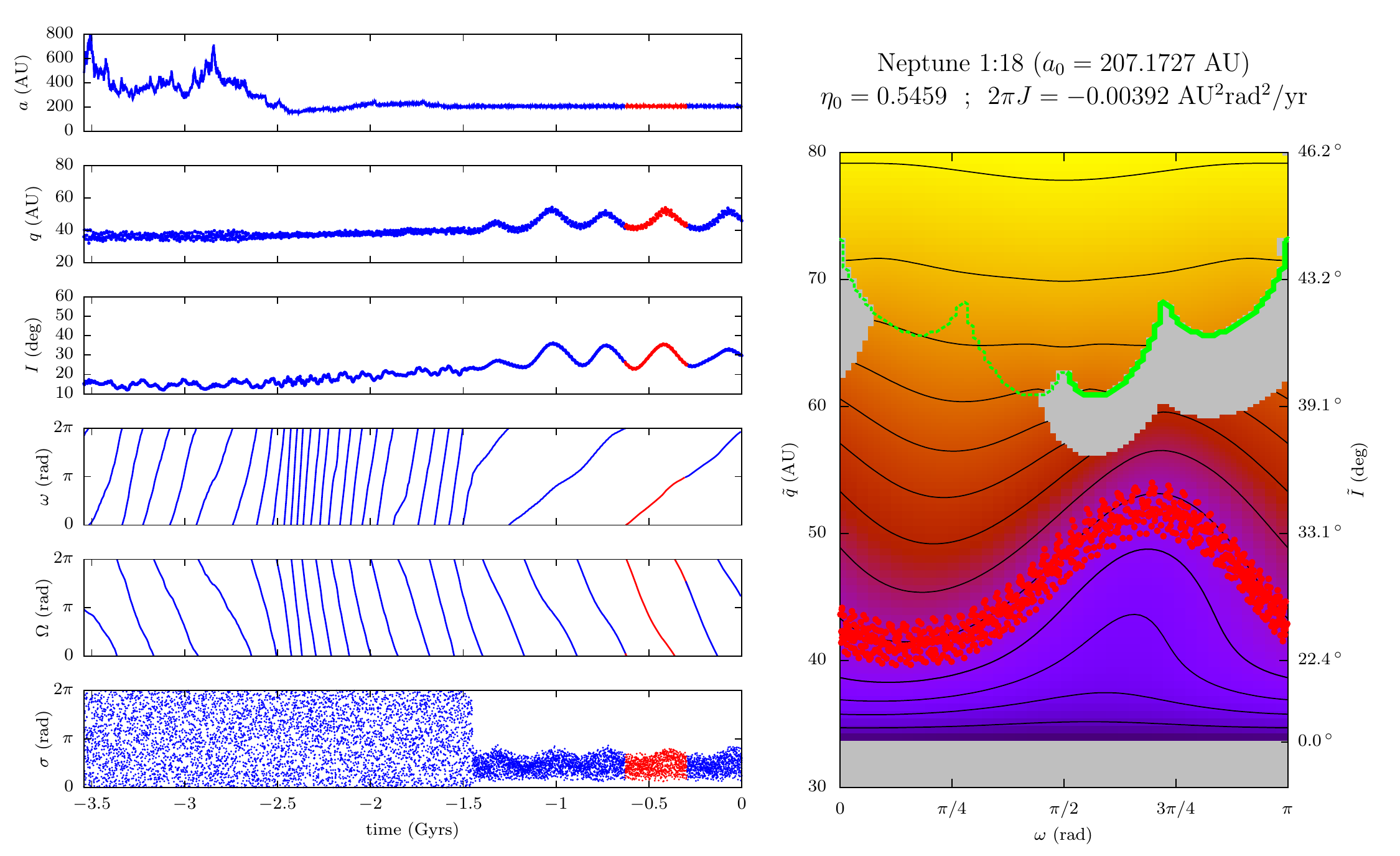}
      \caption{Example of a particle coming from the Oort Cloud and captured in a resonance of type $1\!:\!k$ with Neptune, but with parameters making impossible the transfer to the high-perihelion accumulation zone.}
      \label{fig:integMarc4}
   \end{figure}

   Some examples of orbital evolutions coming from that simulation are shown on Figs.~\ref{fig:integMarc1}-\ref{fig:integMarc4}. The resonant relations with Neptune, necessary to explain the trapping process, are pretty obvious. Figure~\ref{fig:integMarc1} presents a very common example of high-perihelion trapping by means of a resonance of type $1\!:\!k$ with Neptune. The resonant capture is deep (small area $|2\pi J|$) which allows to bring the perihelion to distances where the diffusion of $a$ is not a risk anymore. When the occupied resonance island shrinks and eventually disappears (thick green line), the particle adopts a dynamics with long-term stability: the resonant angle switches smoothly from high-amplitude oscillations inside the single island (beyond the line, in particular when $\omega\approx\pi/2\mod\pi$) to horseshoe oscillations (below the line, in particular when $\omega\approx 0\mod\pi$). This is the second mechanism described in Sect.~\ref{sec:trapping}.
   
   Figure~\ref{fig:integMarc2} shows that the trapping mechanism does not trigger always at the first attempt: in that particular case, the particle switches resonant configurations before finding the perfect entrance. That kind of ``integrable by parts" trajectory was described in Paper I.
   
   On Fig.~\ref{fig:integMarc3}, we see that the reservoir is not absolutely closed, because the slow diffusion of $J$ can still occasion an extra transition \emph{inside} one of the two resonance islands (instead of the more or less grazing horseshoe orbit). In that case, the new secular level curve may lead again the particle to lower perihelion distances, but since every trajectory leaving the discontinuity line goes back to it, the excursion is only temporary. In that specific example, the particle rejoins indeed the reservoir with a perihelion even higher than before. There is however a possibility of definitive escape from the reservoir if the perihelion distance decreases so much as to enter the chaotic region near Neptune, where the overlap of neighbouring resonances can break the quasi-integrable dynamics and trigger the diffusion of $a$. However, that kind of evolution seems to be rather exceptional.
   
   Figure~\ref{fig:integMarc4} presents the case of a resonant capture with inappropriate parameters: the area $|2\pi J|$ is rather high and the parameter $\eta_0$ is too close to the limit of the range of interest (see Fig.~\ref{fig:interest}). Consequently, the particle is unable to reach the discontinuity line and trigger the trapping mechanism. It does not participate to the reservoir as described in Sect.~\ref{sec:trapping}, though \cite{GOMES-etal_2005} would consider it a High Perihelion Scattered Disc Object (HPSDO).
   
\begin{figure}
      \centering
      \includegraphics[width=\textwidth]{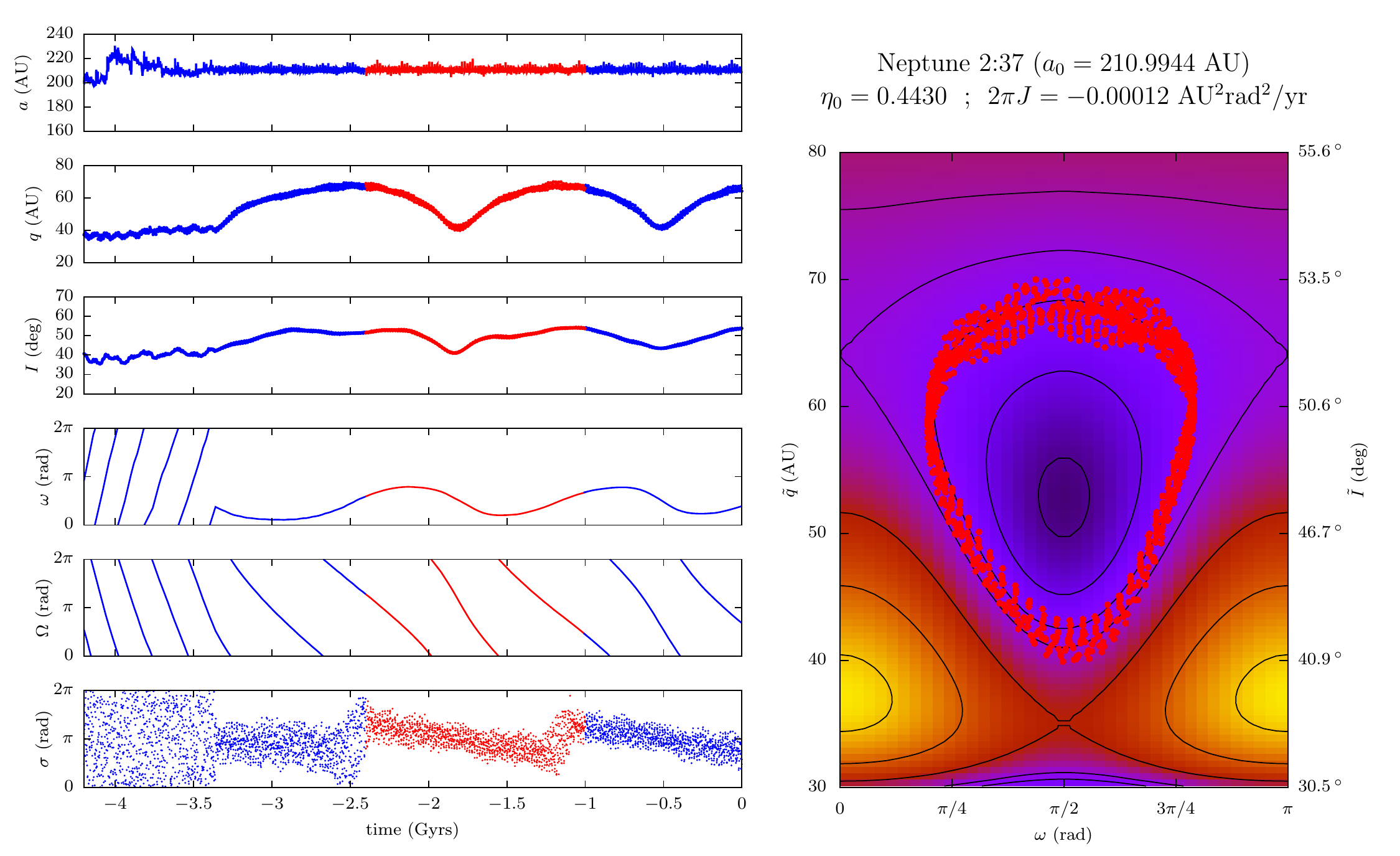}
      \caption{Example of a particle coming from the Oort Cloud and captured in a resonance of type $2\!:\!k$ with Neptune. The perihelion distance reaches high values but the periodic trajectory brings it back toward Neptune. (This numerical integration comes from another sample but the initial conditions are consistent with the distribution given by \cite{FOUCHARD-etal_2016} and used throughout this section).}
      \label{fig:comet83}
   \end{figure}

   Finally, the particle presented on Fig.~\ref{fig:comet83} is a kind of interloper: it is trapped in a resonance of type $2\!:\!k$ which indeed brings it into the range of orbital parameters specific to the reservoir, but its presence there is only temporary since there is no secular discontinuity line able to trigger the trapping mechanism. Such objects participate though to the accumulation (but to a lesser extent), because once trapped in resonance the time spent inside the accumulation zone can be pretty long.
   
   We can estimate the current population contained in the reservoir by the fraction of blue points on Fig.~\ref{fig:reservoir}: we get $42$ objects, to be compared to the $263\,627$ initial conditions coming from the simulation by \cite{FOUCHARD-etal_2016}, that is a fraction of $1.6/10\,000$. With respect to the $10^7$ particles used in that simulation, we get a total fraction of $1/250\,000$.
   
   Note that there is no mention of irreversible trapping in \cite{GOMES-etal_2005} other than the escape out-of-resonance due to Neptune's migration. According to them, every particle driven by a resonant secular dynamics is bound to recover its low-perihelion state after some amount of time. The fact that they did not observe such trappings is probably due to the narrowness of the dynamical path compared to the relatively low number of particles they simulated ($10\,000$ for their largest sample). This could be also an indication that the distribution of the objects coming from the Oort Cloud is more suitable for the trappings to occur.
   
\section{Discussion and conclusion}
   
   \subsection{Resonant secular dynamics beyond Neptune}
   We used the secular model by \cite{SAILLENFEST-etal_2016} to describe the dynamics of transneptunian objects in mean-motion resonance with Neptune. We focused our study on the high-order resonances ($a>100$~AU) and orbits entirely beyond Neptune ($q>30$~AU). That model is characterized by one degree of freedom (linked to the coordinates $\omega$ and $q$) and two parameters ($\eta_0$ and $J$). It proves to represent very accurately the dynamics of a test-particle perturbed by a set of internal circular and coplanar planets (the unaveraged numerical integrations stick astonishingly well to the level curves). Naturally, the dynamics of real transneptunian objects is not so regular, mainly because of the non-zero eccentricities and inclinations of the planets and the secular variation of their orbital elements. These new degrees of freedom have the general effect to add chaotic transitions of $J$ (that is to weaken the resonance captures) and make oscillate the parameter $\eta_0$ (because of the loss of rotational symmetry). In particular, very high-order resonances are unstable for low-inclination orbits in a more realistic planetary system. However, resonances with semi-major axis smaller than $\sim 350$ AU are revealed to be strong enough to capture particles (at least temporarily) and occasion large variations of their orbital elements. The resonant secular model is then very effective to forecast or explain these trajectories. Note that on Fig.~\ref{fig:cutoff}, the orbit remains strongly affected by the resonance up to $q\approx 78$~AU. That example shows that a high perihelion distance and a high semi-major axis (here $a\approx 200$~AU) should not be used as a criterion to state that an object is decoupled from the planets. Indeed, a weak dynamical perturbation does not imply that it cannot have large effects on the orbital elements, especially on a long time-scale.
   
   The notion of ``decoupling" is itself problematic, since of course the planetary perturbations on Solar System objects never vanish completely. We saw that its detection in the sense of ``a weak interaction with the planets letting the orbital elements $(a,e,I)$ quite unchanged" requires an extensive dynamical study. With that definition, Sedna and $2012\text{VP}_{113}$ are indeed decoupled from the giant planets, as both non-resonant and resonant interactions have no other effect for them than making $\omega$ and $\Omega$ circulate. Hence, a ``decoupled" object could be defined as an object located out of the \emph{ranges of interest} (related to large variations of the perihelion distance, see Fig.~\ref{fig:interest}), or, if it is inside one of them, it should be reasonably far from the equilibrium points of the secular Hamiltonian function. Naturally, this implies that the particle is also sufficiently far from the generic Kozai islands (inclination near $63\degree$ or $117\degree$). That definition makes clear that the particles presented in Figs.\ref{fig:integMarc1}-\ref{fig:integMarc3} are very influenced by their resonant relation with Neptune, despite their misleading stable large perihelion distance.
   
   The general geometry of the phase portraits depends very little on the resonance order (that is on the reference semi-major axis $a_0$ of the resonance $k_p\!:\!k$) but is function of its coefficient $k_p$. Indeed, the very same level curves can be obtained for very different values of $a_0$ providing that the parameter $\eta_0$ is modified accordingly (smaller $\eta_0$ for higher $a_0$). In particular, we did not recover the conclusions of \cite{GALLARDO-etal_2012}, stating that large variations of the perihelion distance require a higher inclination for larger semi-major axes (see the graphs A and B of Fig.~\ref{fig:prograde} for a counterexample). The segregation on their Fig.~15 is probably due to the stability of the resonance capture, which indeed requires higher inclinations for larger semi-major axes.
   
   The resonances of type $k_p=1$ are the only ones to present wide regions with two resonance islands. For prograde orbits of this type, there is a clear limit in perihelion distance beyond which there remains only one resonance island. We refer to that limit as the ``secular discontinuity line". The general effect of $\eta_0$ is the same for every kind of resonances (but produces asymmetric features for $1\!:\!k$ resonances). Our conclusions can be summarized as follows:
   \begin{enumerate}
      \item For $\eta_0$ near $0$, $-1$ or $1$ (that is orbits with $I\approx 90\degree$, or circular orbits coplanar to the planetary plane), the phase space presents only circulation zones for $\omega$ with small oscillations of $q$.
      \item For retrograde and prograde orbits, there is always a range of $\eta_0$ at which a libration island appears at $\omega=0\mod\pi$.
      \item For prograde orbits, an additional wide island appears at $\omega=\pi/2\mod\pi$. In the case of a resonance of type $1\!:\!k$, that island is shifted and truncated by the discontinuity line.
      \item In the regions where the resonant part of the secular Hamiltonian weakens (especially at high perihelion distances), the non-resonant libration zones can show up. The interaction between purely resonant features and these ``classic" Kozai islands can lead to very complex geometries. In particular, very wide libration islands can appear in the plane $(\omega,q)$, producing many dynamical paths to high perihelion distances, even from low perihelion and low inclination values.
      \item For moderate values of the semi-major axis (say smaller than $\sim 130$~AU), a very complicated pattern of secular libration islands can appear near the circular orbit (perihelion distances from about $65$~AU to $a$).
      \item For large values of the area $|2\pi J|$, the secular level curves are generally very ``flat" (circulation of $\omega$ with constant $q$). Moreover, a resonance capture at low perihelion distances is very unstable if $|2\pi J|$ is large, and would anyway be broken if $q$ grows too much (separatrix crossing). Consequently, the most interesting trajectories are obtained for deep resonance captures ($J\approx 0$).
   \end{enumerate}
   
   As already stated in previous studies, a resonant interaction with Neptune is found unable to explain the current orbits of Sedna and $2012\text{VP}_{113}$. The majority of the observed objects with $a>100$~AU and $q>30$~AU, though, are located inside the range of parameters that would allow strong variations of their orbital elements in case of mean-motion resonances. The question of whether these objects are indeed in resonance or not is difficult and not studied in this paper.
   
   In addition to bring out many high-perihelion equilibrium points, the resonant secular model highlights a dynamical path from low inclinations and a perihelion near Neptune to a quasi-integrable high-perihelion state with long-term stability. That mechanism is directly linked to the secular discontinuity line, so it is specific to resonances of type $1\!:\!k$. Indeed, the crossing of the secular discontinuity line often triggers a very stable resonant behaviour, where the particle alternates smoothly from high-amplitude oscillations inside the single island (when beyond the line) to horseshoe oscillations (when below it). In that way, there is no more discontinuity and $J$ can be conserved indefinitely. As the final area $|2\pi J|$ is large, $q$ is almost constant and $\omega$ circulates. Thanks to the large value of $q$, the final orbit is stable despite the high-amplitude oscillations of $\sigma$. We estimated the size of that region by the means of the semi-analytical model: it lies approximately in $a\in[100;300]$~AU, $q\in [50;70]$~AU and $I\in [30;50]\degree$ (with circulating angles $\omega$ and $\Omega$). The very long-term stability of that ``reservoir" implies that its population should be increasing since the formation of the planetary system \citep[see also][]{GOMES-etal_2005}. Indeed, it happens to be the end-state of an appreciable number of objects in numerical simulations of transneptunian objects. In particular, the Oort Cloud proved to be a substantial supply of such objects by transferring bodies to the Scattered Disc. According to the initial conditions by \cite{FOUCHARD-etal_2016}, the fraction of Oort Cloud objects transferred to this high-perihelion accumulation zone is of order $1/250\,000$ during the age of the Solar System. This implies that at least $4\times 10^{6}$ objects should currently lie there, considering an initial Oort Cloud population of $10^{12}$ bodies for absolute magnitudes smaller than~11 \citep[consistent with][]{FRANCIS_2005}.
   
   \subsection{Comparison to alternative models}
   We showed that a mean-motion resonance with Neptune can produce very large variations of the perihelion distance and/or confine the argument of perihelion of a transneptunian object. For small perihelion distances (say $<40$~AU), the equilibrium points are all located at $\omega=0\!\mod{\pi}$ whatever the resonance considered (but they are slightly shifted for resonances $1\!:\!k$). However, that mechanism could not explain any preferential location for $\Omega$ or favour $\omega =0$ against $\pi$, because of the nearly circular and coplanar orbits of the giant planets. Such features, if they are really significant in the observed distribution of the transneptunian objects, would require an asymmetric perturbation. It could be an additional eccentric distant planet \citep{BATYGIN-BROWN_2016} or the memory of a captured population from another star \citep{JILKOVA-etal_2015}. In the first case, secular models as developed in \cite{SAILLENFEST-etal_2016} and used in this paper cannot apply to the distant transneptunian objects. In the second case, they can be used to describe the current dynamics of these objects, now that the perturber has gone. Finally, note that a distant super-Earth would turn the far transneptunian region into a very chaotic sea. In particular, the high-perihelion reservoir as described above would not be an accumulation region but part of a continuum of orbital elements. This could be a new observational argument, although it would require an extended sample and unfortunately such objects are very difficult to observe from the Earth.
   
\begin{acknowledgements}
   The authors thank the two anonymous referees for their wise and very stimulating comments. They brought a valuable contribution to the article, allowing a deeper understanding in several parts of the work.
\end{acknowledgements}
   
\bibliographystyle{aps-nameyear}
\bibliography{CM_2016}

\begin{thebibliography}{19}
\ifx \bisbn   \undefined \def \bisbn  #1{ISBN #1}\fi
\ifx \binits  \undefined \def \binits#1{#1} \fi
\ifx \bauthor  \undefined \def \bauthor#1{#1} \fi
\ifx \bjtitle  \undefined \def \bjtitle#1{\textrm{#1}}\fi
\ifx \batitle  \undefined \def \batitle#1{#1} \fi
\ifx \bctitle  \undefined \def \bctitle#1{#1} \fi
\ifx \bvolume  \undefined \def \bvolume#1{\textbf{#1}}\fi
\ifx \byear  \undefined \def \byear#1{#1} \fi
\ifx \bissue  \undefined \def \bissue#1{#1} \fi
\ifx \bfpage  \undefined \def \bfpage#1{#1} \fi
\ifx \blpage  \undefined \def \blpage #1{#1} \fi
\ifx \burl  \undefined \def \burl#1{#1} \fi
\ifx \doiurl  \undefined \def \doiurl#1{#1} \fi
\ifx \betal  \undefined \def \betal{et al.} \fi
\ifx \binstitute  \undefined \def \binstitute#1{#1} \fi
\ifx \beditor  \undefined \def \beditor#1{#1} \fi
\ifx \bpublisher  \undefined \def \bpublisher#1{#1} \fi
\ifx \bbtitle  \undefined \def \bbtitle#1{\textit{#1}} \fi
\ifx \bedition  \undefined \def \bedition#1{#1} \fi
\ifx \bseriesno  \undefined \def \bseriesno#1{#1} \fi
\ifx \blocation  \undefined \def \blocation#1{#1} \fi
\ifx \bsertitle  \undefined \def \bsertitle#1{#1} \fi
\ifx \bsnm \undefined \def \bsnm#1{#1} \fi
\ifx \bsuffix \undefined \def \bsuffix#1{#1} \fi
\ifx \bparticle \undefined \def \bparticle#1{#1} \fi
\ifx \barticle \undefined \def \barticle#1{#1} \fi
\ifx \botherref \undefined \def \botherref #1{#1} \fi
\ifx \url \undefined \def \url#1{#1} \fi
\ifx \bchapter \undefined \def \bchapter#1{#1} \fi
\ifx \bbook \undefined \def \bbook#1{#1} \fi
\ifx \bcomment \undefined \def \bcomment#1{#1} \fi
\ifx \oauthor \undefined \def \oauthor#1{#1} \fi
\ifx \citeauthoryear \undefined \def \citeauthoryear#1{#1} \fi
\ifx \texttildelow  \undefined \def \texttildelow{\symbol{126}} \fi
\def \endbibitem {}
\ifx \bconflocation  \undefined \def \bconflocation#1{#1} \fi

\bibitem[\protect\citeauthoryear{{Batygin} and
  {Brown}}{2016}]{BATYGIN-BROWN_2016}
\begin{barticle}
\bauthor{\binits{K.} \bsnm{{Batygin}}},
\bauthor{\binits{M.E.} \bsnm{{Brown}}},
\batitle{{Evidence for a Distant Giant Planet in the Solar System}}.
\bjtitle{\aj}
\bvolume{151},
\bfpage{22}
(\byear{2016})
\end{barticle}
\endbibitem

\bibitem[\protect\citeauthoryear{{Duncan} et~al.}{1995}]{DUNCAN-etal_1995}
\begin{barticle}
\bauthor{\binits{M.J.} \bsnm{{Duncan}}},
\bauthor{\binits{H.F.} \bsnm{{Levison}}},
\bauthor{\binits{S.M.} \bsnm{{Budd}}},
\batitle{{The Dynamical Structure of the Kuiper Belt}}.
\bjtitle{\aj}
\bvolume{110},
\bfpage{3073}
(\byear{1995})
\end{barticle}
\endbibitem

\bibitem[\protect\citeauthoryear{{Fouchard} et~al.}{2016}]{FOUCHARD-etal_2016}
\begin{botherref}
\oauthor{\binits{M.} \bsnm{{Fouchard}}},
\oauthor{\binits{H.} \bsnm{{Rickman}}},
\oauthor{\binits{C.} \bsnm{{Froeschl{\'e}}}},
\oauthor{\binits{G.B.} \bsnm{{Valsecchi}}},
{On the present shape of the Oort cloud and the flux of ``new" comets}
\textbf{{ -- accepted for publication in Icarus}}
(2016)
\end{botherref}
\endbibitem

\bibitem[\protect\citeauthoryear{{Francis}}{2005}]{FRANCIS_2005}
\begin{barticle}
\bauthor{\binits{P.J.} \bsnm{{Francis}}},
\batitle{{The Demographics of Long-Period Comets}}.
\bjtitle{\apj}
\bvolume{635},
\bfpage{1348}--\blpage{1361}
(\byear{2005})
\end{barticle}
\endbibitem

\bibitem[\protect\citeauthoryear{{Gallardo} et~al.}{2012}]{GALLARDO-etal_2012}
\begin{barticle}
\bauthor{\binits{T.} \bsnm{{Gallardo}}},
\bauthor{\binits{G.} \bsnm{{Hugo}}},
\bauthor{\binits{P.} \bsnm{{Pais}}},
\batitle{{Survey of Kozai dynamics beyond Neptune}}.
\bjtitle{\ica}
\bvolume{220},
\bfpage{392}--\blpage{403}
(\byear{2012})
\end{barticle}
\endbibitem

\bibitem[\protect\citeauthoryear{{Gladman} et~al.}{2002}]{GLADMAN-etal_2002}
\begin{barticle}
\bauthor{\binits{B.} \bsnm{{Gladman}}},
\bauthor{\binits{M.} \bsnm{{Holman}}},
\bauthor{\binits{T.} \bsnm{{Grav}}},
\bauthor{\binits{J.} \bsnm{{Kavelaars}}},
\bauthor{\binits{P.} \bsnm{{Nicholson}}},
\bauthor{\binits{K.} \bsnm{{Aksnes}}},
\bauthor{\binits{J.-M.} \bsnm{{Petit}}},
\batitle{{Evidence for an Extended Scattered Disk}}.
\bjtitle{\ica}
\bvolume{157},
\bfpage{269}--\blpage{279}
(\byear{2002})
\end{barticle}
\endbibitem

\bibitem[\protect\citeauthoryear{{Gomes}}{2011}]{GOMES_2011}
\begin{barticle}
\bauthor{\binits{R.S.} \bsnm{{Gomes}}},
\batitle{{The origin of TNO 2004 XR $_{190}$ as a primordial scattered
  object}}.
\bjtitle{\ica}
\bvolume{215},
\bfpage{661}--\blpage{668}
(\byear{2011})
\end{barticle}
\endbibitem

\bibitem[\protect\citeauthoryear{{Gomes} et~al.}{2005}]{GOMES-etal_2005}
\begin{barticle}
\bauthor{\binits{R.S.} \bsnm{{Gomes}}},
\bauthor{\binits{T.} \bsnm{{Gallardo}}},
\bauthor{\binits{J.A.} \bsnm{{Fern{\'a}ndez}}},
\bauthor{\binits{A.} \bsnm{{Brunini}}},
\batitle{{On The Origin of The High-Perihelion Scattered Disk: The Role of The
  Kozai Mechanism And Mean Motion Resonances}}.
\bjtitle{\cmda}
\bvolume{91},
\bfpage{109}--\blpage{129}
(\byear{2005})
\end{barticle}
\endbibitem

\bibitem[\protect\citeauthoryear{{H{\'e}non}}{1982}]{HENON_1982}
\begin{barticle}
\bauthor{\binits{M.} \bsnm{{H{\'e}non}}},
\batitle{{On the numerical computation of Poincar{\'e} maps}}.
\bjtitle{Physica D Nonlinear Phenomena}
\bvolume{5},
\bfpage{412}--\blpage{414}
(\byear{1982})
\end{barticle}
\endbibitem

\bibitem[\protect\citeauthoryear{{Henrard}}{1993}]{HENRARD_1993}
\begin{bbook}
\bauthor{\binits{J.} \bsnm{{Henrard}}},
\bbtitle{{Dynamics Reported -- Expositions in Dynamical Systems}},
vol. \bseriesno{2}
(\bpublisher{{{Springer Berlin Heidelberg}}}, \byear{1993}),
pp. \bfpage{117}--\blpage{235}
\end{bbook}
\endbibitem

\bibitem[\protect\citeauthoryear{{Holman} and
  {Wisdom}}{1993}]{HOLMAN-WISDOM_1993}
\begin{barticle}
\bauthor{\binits{M.J.} \bsnm{{Holman}}},
\bauthor{\binits{J.} \bsnm{{Wisdom}}},
\batitle{{Dynamical stability in the outer solar system and the delivery of
  short period comets}}.
\bjtitle{\aj}
\bvolume{105},
\bfpage{1987}--\blpage{1999}
(\byear{1993})
\end{barticle}
\endbibitem

\bibitem[\protect\citeauthoryear{{J{\'{\i}}lkov{\'a}}
  et~al.}{2015}]{JILKOVA-etal_2015}
\begin{barticle}
\bauthor{\binits{L.} \bsnm{{J{\'{\i}}lkov{\'a}}}},
\bauthor{\binits{S.} \bsnm{{Portegies Zwart}}},
\bauthor{\binits{T.} \bsnm{{Pijloo}}},
\bauthor{\binits{M.} \bsnm{{Hammer}}},
\batitle{{How Sedna and family were captured in a close encounter with a solar
  sibling}}.
\bjtitle{\mnras}
\bvolume{453},
\bfpage{3157}--\blpage{3162}
(\byear{2015})
\end{barticle}
\endbibitem

\bibitem[\protect\citeauthoryear{{Kozai}}{1962}]{KOZAI_1962}
\begin{barticle}
\bauthor{\binits{Y.} \bsnm{{Kozai}}},
\batitle{{Secular perturbations of asteroids with high inclination and
  eccentricity}}.
\bjtitle{\aj}
\bvolume{67},
\bfpage{591}
(\byear{1962})
\end{barticle}
\endbibitem

\bibitem[\protect\citeauthoryear{{Kozai}}{1985}]{KOZAI_1985}
\begin{barticle}
\bauthor{\binits{Y.} \bsnm{{Kozai}}},
\batitle{{Secular perturbations of resonant asteroids}}.
\bjtitle{\celmec}
\bvolume{36},
\bfpage{47}--\blpage{69}
(\byear{1985})
\end{barticle}
\endbibitem

\bibitem[\protect\citeauthoryear{{Laskar}}{1990}]{LASKAR_1990}
\begin{barticle}
\bauthor{\binits{J.} \bsnm{{Laskar}}},
\batitle{{The chaotic motion of the solar system - A numerical estimate of the
  size of the chaotic zones}}.
\bjtitle{\ica}
\bvolume{88},
\bfpage{266}--\blpage{291}
(\byear{1990})
\end{barticle}
\endbibitem

\bibitem[\protect\citeauthoryear{{Milani} and
  {Baccili}}{1998}]{MILANI-BACCILI_1998}
\begin{barticle}
\bauthor{\binits{A.} \bsnm{{Milani}}},
\bauthor{\binits{S.} \bsnm{{Baccili}}},
\batitle{{Dynamics of Earth-crossing asteroids: the protected Toro orbits}}.
\bjtitle{\cmda}
\bvolume{71},
\bfpage{35}--\blpage{53}
(\byear{1998})
\end{barticle}
\endbibitem

\bibitem[\protect\citeauthoryear{{Saillenfest}
  et~al.}{2016}]{SAILLENFEST-etal_2016}
\begin{barticle}
\bauthor{\binits{M.} \bsnm{{Saillenfest}}},
\bauthor{\binits{M.} \bsnm{{Fouchard}}},
\bauthor{\binits{G.} \bsnm{{Tommei}}},
\bauthor{\binits{G.B.} \bsnm{{Valsecchi}}},
\batitle{{Long term dynamics beyond Neptune: secular models to study the
  regular motions}}.
\bjtitle{\cmda}
\bvolume{126},
\bfpage{369}--\blpage{403}
(\byear{2016})
\end{barticle}
\endbibitem

\bibitem[\protect\citeauthoryear{{Sheppard} et~al.}{2016}]{SHEPPARD-etal_2016}
\begin{barticle}
\bauthor{\binits{S.S.} \bsnm{{Sheppard}}},
\bauthor{\binits{C.} \bsnm{{Trujillo}}},
\bauthor{\binits{D.J.} \bsnm{{Tholen}}},
\batitle{{Beyond the Kuiper Belt Edge: New High Perihelion Trans-Neptunian
  Objects with Moderate Semimajor Axes and Eccentricities}}.
\bjtitle{\apjl}
\bvolume{825},
\bfpage{13}
(\byear{2016})
\end{barticle}
\endbibitem

\bibitem[\protect\citeauthoryear{{Thomas} and
  {Morbidelli}}{1996}]{THOMAS-MORBIDELLI_1996}
\begin{barticle}
\bauthor{\binits{F.} \bsnm{{Thomas}}},
\bauthor{\binits{A.} \bsnm{{Morbidelli}}},
\batitle{{The Kozai Resonance in the Outer Solar System and the Dynamics of
  Long-Period Comets}}.
\bjtitle{{\cmda}}
\bvolume{64},
\bfpage{209}--\blpage{229}
(\byear{1996})
\end{barticle}
\endbibitem

\end{thebibliography}

\end{document}